# Time–frequency distributions of optical sources (I): directly modulated semiconductor lasers

**José Capmany***

*ITEAM Research Institute, Universitat Politècnica de València, Camino de Vera s/n, 46022 Valencia, Spain*

*jcapmany@iteam.upv.es

**Abstract:** We develop a time–frequency framework for the dynamics of directly modulated semiconductor lasers in which the optical field obtained from the single-mode rate equations is represented by distributions of Cohen's class. Closed-form small-signal solutions are derived for RC-filtered square current pulses and for sinusoidal modulation, including the exact chirp-to-power ratio with gain compression and spontaneous emission and a line expansion of the Wigner distribution of the sinusoidally modulated field. Kernels satisfying two known conditions return the laser chirp as the first conditional moment; the transient chirp equals $\alpha(1 + \varepsilon S)$ times Cohen's instantaneous bandwidth; and the rms spectral width of a chirped pulse obeys an exact budget that reduces to $B^2 = (1 + \alpha^2)B_{TL}^2 + \sigma_{ad}^2$ for weak compression. A four-term Koch–Linke regression on the moments removes the bias of the standard form, whereas spectrogram moments bias $\alpha$ by up to 17%; the sensitivity to noise and receiver bandwidth is quantified. For two-tone modulation, a Volterra expansion of the rate equations gives closed-form second- and third-order intermodulation, and a Langevin analysis gives the relative intensity noise; together they yield the spurious-free dynamic ranges in 1 Hz, e.g., 75.7 dB·Hz$^{1/2}$ (second order) and 109.7 dB·Hz$^{2/3}$ (third order) at 2 GHz. Results for a 1550 nm DFB laser, with 1310 nm DFB and 850 nm VCSEL comparisons, validate the closed-form results against numerical integration.

## 1. Introduction

Direct current modulation remains the simplest and most cost-effective way of imprinting information on the output of a semiconductor laser, and directly modulated lasers (DMLs) are ubiquitous in short-reach optical links, radio-over-fiber systems and pulsed sources [1–3]. Their dynamics are governed by the coupled evolution of the carrier and photon populations, whose interplay produces relaxation oscillations, a finite modulation bandwidth and harmonic and intermodulation distortion [4–6]. Because the refractive index of the active region depends on the carrier density, intensity modulation is necessarily accompanied by frequency modulation, quantified by the linewidth-enhancement factor $\alpha$ [7]. The resulting frequency chirp consists of a transient component, proportional to the logarithmic derivative of the optical power, and an adiabatic component, proportional to the power itself and originating from nonlinear gain compression [8–10]. The two components have very different consequences: transient chirp broadens the optical spectrum and interacts with fiber dispersion on the time scale of the transitions, whereas adiabatic chirp produces a power-dependent carrier frequency that can be exploited, for instance, in chirp-managed lasers [11].

The standard descriptions of DML chirp are either purely temporal (the instantaneous frequency deviation as a function of time) or purely spectral (the optical spectrum and its modulation sidebands [12,13]). Neither representation shows simultaneously *when* a given frequency is emitted and *how much* energy is associated with it, and neither exposes directly the geometric relation between the amplitude and the phase of the field that the $\alpha$-factor establishes. Joint time–frequency distributions do exactly this. The Wigner–Ville distribution (WVD) [14,15] and, more generally, Cohen's class of quadratic distributions [16–19] provide energy densities in the time–frequency plane whose marginals and moments have precise physical meaning [20]. In optics the Wigner distribution has a long history as a phase-space representation of beams and pulses [21,22]; the chronocyclic representation of ultrashort pulses [23] and spectrographic pulse-measurement techniques such as frequency-resolved optical gating (FROG) [24,25] are its best-known applications to temporal optics, and spectrograms have been used to characterize telecommunication pulses and gain-switched diode lasers [26,27]. Marcenac and Carroll proposed the Wigner distribution as a device-independent description of the spectral dynamics of laser diodes and illustrated it with calculated time-resolved spectra of a gain-switched laser [28].

In this paper we develop that idea into a systematic and quantitative framework, summarized in Fig. 1. The field of a single-mode laser, obtained from the rate equations under RC-filtered square-pulse and sinusoidal current modulation, is analyzed with several members of Cohen's class. The contributions are the following. (i) We derive an exact decomposition of the chirp into transient, adiabatic and spontaneous-emission terms, and closed-form small-signal solutions for the power and chirp under RC-filtered rectangular pulses and under sinusoidal modulation, including the exact chirp-to-power ratio with gain compression and spontaneous emission. (ii) We obtain a closed-form line expansion of the Wigner distribution of the sinusoidally modulated field and a criterion, based on the frequency-modulation index, for the visibility of the instantaneous-frequency ridge. (iii) Using the known kernel conditions under which the first conditional moment of a distribution equals the instantaneous frequency [17,20], we show that the transient chirp equals $\alpha(1+\varepsilon S)$ times Cohen's instantaneous bandwidth and derive an exact rms-bandwidth budget for chirped laser pulses. (iv) We show that the Koch–Linke analysis used with time-resolved chirp measurements [29,30] can be carried out on the moments of the distribution, that retaining the compression factor in the transient term removes the bias of the standard three-term form, that spectrogram moments are intrinsically biased, and we quantify the sensitivity of the estimators to noise and receiver bandwidth. (v) For two-tone modulation we derive the Volterra kernels of the rate equations to third order and a Langevin expression for the RIN. From them we obtain the intercept points and the second- and third-order SFDR of a directly modulated analog link in a 1 Hz bandwidth, together with the intermodulation of the frequency modulation. The closed-form results are validated against direct numerical integration of the rate equations for a 1550 nm distributed-feedback (DFB) laser, and the framework is applied comparatively to a 1310 nm DFB laser and an 850 nm vertical-cavity surface-emitting laser (VCSEL).

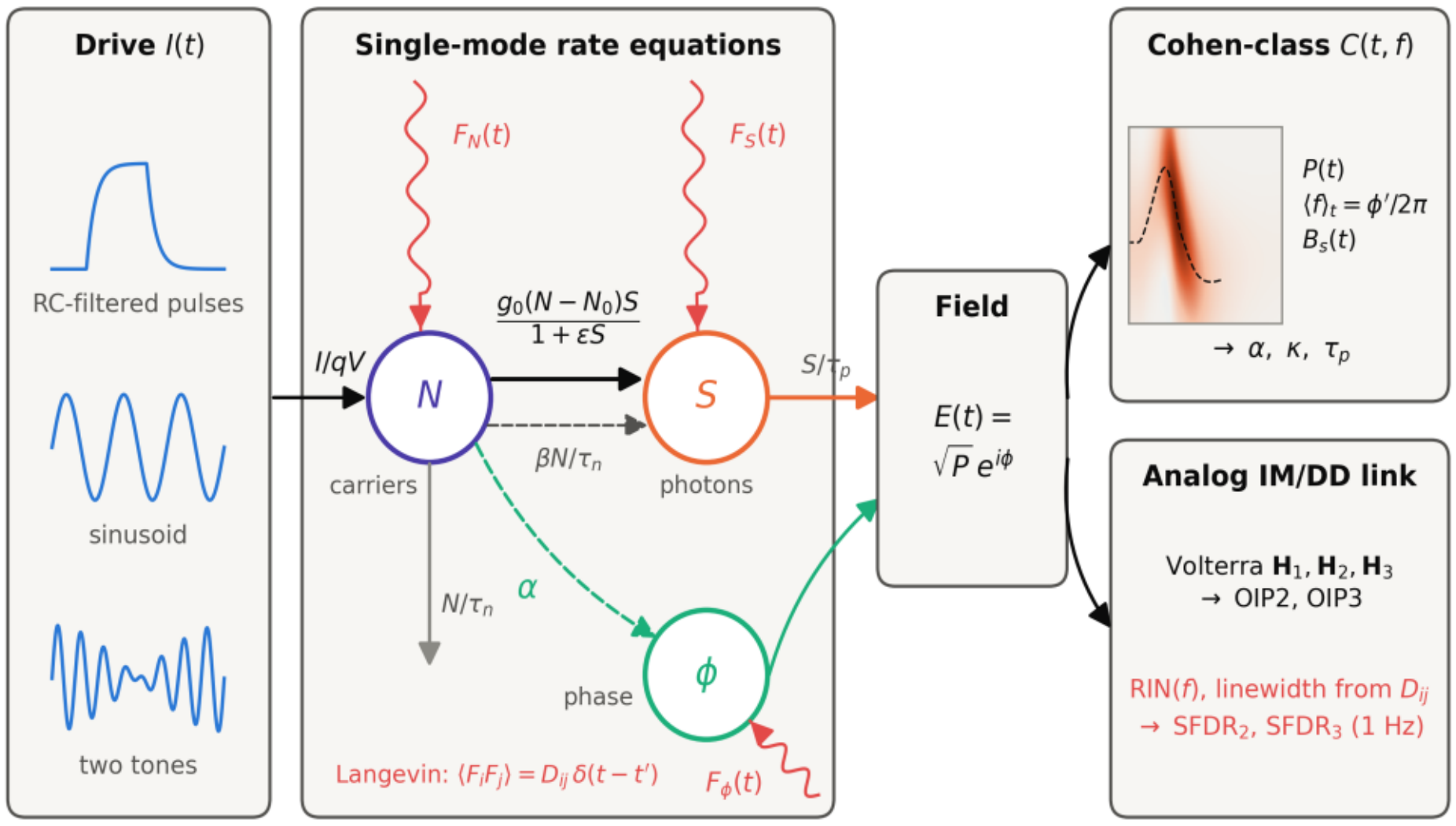

**Fig. 1.** Framework of this work. RC-filtered pulses, sinusoidal or two-tone currents drive the single-mode rate equations, in which carriers ($N$) and photons ($S$) are coupled by stimulated and spontaneous emission and perturbed by the Langevin sources $F_N$ and $F_S$; the carrier density sets the optical phase $\phi$ through the $\alpha$-factor, and the phase is driven by the Langevin source $F_\phi$. The complex field $E(t)$ is analyzed along two branches: Cohen-class time–frequency distributions, whose marginals and conditional moments yield the power, the chirp, the instantaneous bandwidth and the laser parameters (inset: PWVD of an RC-filtered pulse); and an analog intensity-modulation/direct-detection link, where Volterra kernels give the intercept points and the Langevin diffusion coefficients $D_{ij}$ give the RIN, the linewidth and the SFDRs.

## 2. Rate-equation model and chirp decomposition

### *2.1 Single-mode rate equations*

We consider a single-longitudinal-mode laser described by the carrier density $N(t)$, the photon density $S(t)$ and the optical phase $\phi(t)$ [1,2]:

$$\frac{dN}{dt} = \frac{I(t)}{qV} - \frac{N}{\tau_n} - \frac{g_0(N-N_0)S}{1+\varepsilon S} + F_N(t), \tag{1}$$

$$\frac{dS}{dt} = \frac{\Gamma g_0(N-N_0)S}{1+\varepsilon S} - \frac{S}{\tau_p} + \frac{\Gamma\beta N}{\tau_n} + F_S(t), \tag{2}$$

$$\frac{d\phi}{dt} = \frac{\alpha}{2}\left[\Gamma g_0(N-N_0) - \frac{1}{\tau_p}\right] + F_\phi(t), \tag{3}$$

where $I(t)$ is the injection current, $q$ the electron charge, $V$ the active volume, $\tau_n$ the carrier lifetime, $g_0 = v_g a$ the differential gain coefficient, $N_0$ the transparency density, $\varepsilon$ the gain-compression factor, $\Gamma$ the confinement factor, $\tau_p$ the photon lifetime and $\beta$ the spontaneous-emission coupling factor. $F_N(t)$, $F_S(t)$ and $F_\phi(t)$ are zero-mean Langevin noise sources, whose correlations are given in Section 6; they are set to zero in the deterministic simulations and enter the relative-intensity-noise analysis and the stochastic validation of Section 6. As is customary, the phase equation Eq. (3) uses the uncompressed modal gain: the carrier-induced refractive-index change follows $N$, whereas the nonlinear gain suppression (spectral hole burning and carrier heating) is taken as index-neutral [8,10]. Frequencies are referred to the emission frequency $\nu_{th}$ of the lasing mode at $N = N_{th}$: the physical field is proportional to $E(t)e^{i2\pi\nu_{th}t}$, so that a positive frequency offset corresponds to a blue shift, and the output of the model is the complex baseband envelope

$$E(t) = \sqrt{P(t)}\, e^{i\phi(t)}, \qquad P(t) = P_S\, S(t), \qquad P_S = \frac{\eta V h\nu}{\Gamma \tau_p}, \tag{4}$$

with $\eta$ the fraction of the internal photon loss emitted through the output facet and $h\nu$ the photon energy ($h$ is Planck's constant; the lag window introduced in Section 7 is denoted $h(\tau)$). The threshold carrier density and current are $N_{th} = N_0 + \left(\Gamma g_0 \tau_p\right)^{-1}$ and $I_{th} = qVN_{th}/\tau_n$.

The parameters of the three lasers studied are listed in Table 1. The 1550 nm DFB laser is the reference device of this work. Its parameters are representative of the values extracted for InGaAsP multiple-quantum-well DFB lasers in [29,30]; the 1310 nm DFB and 850 nm VCSEL sets represent, respectively, a faster laser with a lower $\alpha$ and a small-volume, low-threshold device of the type reviewed in [31,32]. The VCSEL is treated as single-mode, i.e., transverse-mode dynamics are not considered. All three sets are bulk-equivalent rate-equation parameters and are not fits to specific commercial devices.

**Table 1. Rate-equation parameters of the three lasers. DFB-1550 is the reference device; derived quantities are evaluated from Eqs. (1)–(7).**

| Symbol | Description | DFB-1550 | DFB-1310 | VCSEL-850 | Unit |
|---|---|---|---|---|---|
| $\lambda$ | wavelength | 1550 | 1310 | 850 | nm |
| $V$ | active volume | $1.0 \times 10^{-16}$ | $6.0 \times 10^{-17}$ | $6.0 \times 10^{-19}$ | m³ |
| $\Gamma$ | confinement factor | 0.30 | 0.25 | 0.04 | – |
| $g_0$ | differential gain coefficient $v_g a$ | $2.08 \times 10^{-12}$ | $3.3 \times 10^{-12}$ | $8.0 \times 10^{-12}$ | m³ s⁻¹ |
| $N_0$ | transparency density | $1.0 \times 10^{24}$ | $0.9 \times 10^{24}$ | $1.8 \times 10^{24}$ | m⁻³ |
| $\tau_n$ | carrier lifetime | 1.0 | 0.7 | 1.0 | ns |
| $\tau_p$ | photon lifetime | 2.0 | 1.6 | 2.5 | ps |
| $\varepsilon$ | gain-compression factor | $1.5 \times 10^{-23}$ | $1.0 \times 10^{-23}$ | $1.0 \times 10^{-23}$ | m³ |
| $\beta$ | spontaneous-emission coupling | $10^{-4}$ | $10^{-4}$ | $10^{-3}$ | – |
| $\alpha$ | linewidth-enhancement factor | 3.0 | 2.2 | 3.5 | – |
| $\eta$ | output coupling fraction | 0.3 | 0.3 | 0.5 | – |
| $N_{th}$ | threshold density (derived) | $1.80 \times 10^{24}$ | $1.66 \times 10^{24}$ | $3.05 \times 10^{24}$ | m⁻³ |
| $I_{th}$ | threshold current (derived) | 28.9 | 22.8 | 0.293 | mA |
| $\alpha\kappa/4\pi$ | adiabatic chirp coefficient (derived) | 0.279 | 0.160 | 1.589 | GHz mW⁻¹ |

### *2.2 Exact chirp decomposition*

The instantaneous frequency deviation (chirp) is $\Delta\nu(t) = \dot{\phi}/2\pi$. Eliminating $\Gamma g_0(N - N_0)$ between Eqs. (2) and (3) gives, without approximation,

$$\Delta\nu(t) = \frac{\alpha}{4\pi}\left[(1+\varepsilon S)\frac{d\ln S}{dt} + \frac{\varepsilon S}{\tau_p} - (1+\varepsilon S)\frac{\Gamma\beta N}{\tau_n S}\right]. \tag{5}$$

The three terms in brackets are, respectively, the transient, adiabatic and spontaneous-emission contributions to the chirp. The spontaneous term is significant only when the photon density is very small, i.e., when the laser is close to being switched off; in that regime the deterministic phase of Eq. (3) loses physical meaning, since the field is dominated by amplified spontaneous emission, and we accordingly omit the chirp curves wherever the power is below 2% of its peak in the figures Neglecting $\beta$ and using $P = P_S S$ and $d\ln S/dt = d\ln P/dt$, Eq. (5) becomes

$$\Delta\nu(t) = \frac{\alpha}{4\pi}\left[\left(1+\tau_p \kappa P\right)\frac{d\ln P}{dt} + \kappa P\right], \tag{6}$$

$$\kappa = \frac{\varepsilon}{\tau_p P_S}, \qquad \varepsilon S = \tau_p \kappa P, \tag{7}$$

which reduces to the Koch–Linke formula $\Delta\nu = (\alpha/4\pi)[d\ln P/dt + \kappa P]$ [8] when $\varepsilon S \ll 1$. The factor $(1 + \tau_p \kappa P)$ multiplying the transient term is dropped in the Koch–Linke approximation; we retain it because, as shown in Section 7.4, it is essential for an unbiased extraction of $\alpha$ from large-signal data.

## 3. Small-signal analysis

### 3.1 Linearized equations

Let $(\overline{N}, \overline{S})$ be the steady state at bias current $I_b$, and $\delta N$, $\delta S$, $\delta I$ small deviations. Writing $G(N,S) = g_0(N - N_0)/(1 + \varepsilon S)$, $\overline{G} = G(\overline{N}, \overline{S})$ and $G_N = g_0/(1 + \varepsilon\overline{S})$, linearization of Eqs. (1) and (2) gives

$$\frac{d}{dt}\begin{bmatrix}\delta N\\ \delta S\end{bmatrix} = \begin{bmatrix}a_{11} & a_{12}\\ a_{21} & a_{22}\end{bmatrix}\begin{bmatrix}\delta N\\ \delta S\end{bmatrix} + \begin{bmatrix}\delta I/qV\\ 0\end{bmatrix}, \tag{8}$$

with

$$a_{11} = -\frac{1}{\tau_n} - G_N\overline{S}, \qquad a_{12} = -\frac{\overline{G}}{1+\varepsilon\overline{S}}, \tag{9}$$

$$a_{21} = \Gamma G_N \overline{S} + \frac{\Gamma\beta}{\tau_n}, \qquad a_{22} = -\frac{\Gamma\beta\overline{N}}{\tau_n\overline{S}} - \frac{\Gamma\varepsilon\overline{G}\,\overline{S}}{1+\varepsilon\overline{S}}, \tag{10}$$

where the steady-state condition $\Gamma\overline{G} - 1/\tau_p = -\Gamma\beta\overline{N}/(\tau_n\overline{S})$ has been used in $a_{22}$. The characteristic polynomial $Q(s) = s^2 + \gamma s + \omega_R^2$ defines the relaxation-oscillation frequency and damping rate

$$\omega_R^2 = a_{11}a_{22} - a_{12}a_{21}, \qquad \gamma = -(a_{11} + a_{22}), \tag{11}$$

and the intensity-modulation (IM) transfer function is

$$\frac{\delta S(s)}{\delta I(s)} = \frac{a_{21}}{qV}\frac{1}{Q(s)}, \qquad H(s) = \frac{\omega_R^2}{Q(s)}. \tag{12}$$

Equations (9)–(12) contain gain compression and spontaneous emission exactly; the familiar expressions [2,4] are recovered for $\beta \to 0$ and $\varepsilon\overline{S} \ll 1$.

### 3.2 Exact small-signal chirp

Linearizing Eq. (3) gives $\delta\nu = (\alpha/4\pi)\Gamma g_0\,\delta N$, and the second row of Eq. (8) gives $\delta N = (s - a_{22})\delta S/a_{21}$ in the Laplace domain. Defining the relative power modulation $p = \delta P/\overline{P} = \delta S/\overline{S}$,

$$\delta\nu(s) = \frac{\alpha}{4\pi}\zeta(s + \omega_c)p(s), \qquad \zeta = \frac{\Gamma g_0\overline{S}}{a_{21}}, \qquad \omega_c = -a_{22}. \tag{13}$$

In the time domain this is $\delta\nu(t) = (\alpha/4\pi)\zeta[\dot{p} + \omega_c p]$, the exact small-signal form of Eq. (6): the corner frequency $f_c = \omega_c/2\pi$ separating adiabatic and transient behavior is $f_c \simeq [\kappa\overline{P}/(1 + \varepsilon\overline{S}) + \Gamma\beta\overline{N}/(\tau_n\overline{S})]/2\pi$, and $\zeta = (1 + \varepsilon\overline{S})$ up to terms in $\beta$. The chirp-to-power ratio (CPR) at modulation frequency $f$ is $\mathrm{CPR}(f) = (\alpha/4\pi)\zeta(i2\pi f + \omega_c)$; its magnitude tends to the adiabatic value $(\alpha/4\pi)\zeta\omega_c$ below $f_c$ and grows as $(\alpha/2)\zeta f$ above it, while its phase rises from 0 to 90° [8,9]. The small-signal quantities of the reference laser at three bias currents are listed in Table 2.

**Table 2. Small-signal quantities of the DFB-1550 laser from Eqs. (8)–(13).**

| Quantity | $1.5I_{th}$ | $2.5I_{th}$ | $3.5I_{th}$ | Unit |
|---|---|---|---|---|
| output power $\overline{P}$ | 3.44 | 10.31 | 17.19 | mW |
| $\varepsilon\overline{S}$ | 0.008 | 0.024 | 0.040 | – |
| $f_R = \omega_R/2\pi$ | 3.76 | 6.46 | 8.27 | GHz |
| $\gamma/2\pi$ | 0.99 | 2.56 | 4.09 | GHz |
| IM 3-dB bandwidth | 5.77 | 9.75 | 12.29 | GHz |
| corner frequency $f_c$ | 0.65 | 1.88 | 3.08 | GHz |
| $\zeta$ | 1.008 | 1.024 | 1.040 | – |
| $\lvert\mathrm{CPR}(0)\rvert/\overline{P}$ | 0.286 | 0.280 | 0.280 | GHz mW⁻¹ |

## 4. Closed-form solutions for RC-filtered square-pulse modulation

### 4.1 Drive waveform

Digital modulation is modeled as a rectangular (non-return-to-zero) current waveform passed through a first-order low-pass filter of time constant $\tau_{RC}$, which represents the combined bandwidth limitation of the driver, the bond wires and the laser parasitics. Denoting by $t_k$ the bit transitions and by $e_k = \pm 1$ their signs, the exact solution of $\tau_{RC}\, dI/dt = I_{sq}(t) - I$ is

$$I(t) = I_0 + (I_1 - I_0) \sum_k e_k\, u(t - t_k)\big[1 - e^{-(t-t_k)/\tau_{RC}}\big], \tag{14}$$

where $I_0$ and $I_1$ are the currents of the "0" and "1" levels and $u$ is the unit step. A single pulse of width $T$ starting at $t_0$ has $e_0 = +1$ at $t_0$ and $e_1 = -1$ at $t_0 + T$. The 10–90% rise time of the current is $2.2\tau_{RC}$. Gaussian or raised-cosine pulse shapes are not used in this work.

### 4.2 Small-signal response

For a pulse of height $\delta I$ on the bias $I_b$, the relative power modulation is the response of the cascade $\left[a_{21}/\left(qV\overline{S}\right)\right]\,[1/Q(s)]\,[1/(1+s\tau_{RC})]$ to a rectangular pulse. The step response of $1/[(1+s\tau_{RC})Q(s)]$ is obtained by partial fractions at the poles $s = 0$, $s = -1/\tau_{RC}$ and $s_{1,2} = -\gamma/2 \pm i\omega_d$, with $\omega_d = (\omega_R^2 - \gamma^2/4)^{1/2}$:

$$g(t) = \frac{1}{\omega_R^2} - \frac{e^{-t/\tau_{RC}}}{Q(-1/\tau_{RC})} + 2\,\mathrm{Re}\left[\frac{e^{s_1 t}}{2i\omega_d\, s_1(1+\tau_{RC}s_1)}\right], \quad t \ge 0, \tag{15}$$

with $g(t) = 0$ for $t < 0$. Equation (15) assumes distinct poles, i.e., an underdamped response ($\omega_R > \gamma/2$), which holds at all bias points considered here. The power and chirp responses then follow from linearity and Eq. (13):

$$p(t) = \frac{a_{21}\,\delta I}{qV\overline{S}}\,[g(t-t_0) - g(t-t_0-T)], \tag{16}$$

$$\delta\nu(t) = \frac{\alpha}{4\pi}\,\zeta\left[\frac{dp}{dt} + \omega_c\, p(t)\right]. \tag{17}$$

Equations (15)–(17) describe the complete small-signal dynamics: the RC filter adds a real pole that slows the leading and trailing edges and reduces the excitation of the relaxation oscillation, while the transient chirp, proportional to $\dot{p}$, is reduced in proportion. For $\tau_{RC} \to 0$ the classical second-order step response is recovered.

### 4.3 Turn-on delay below threshold

When the "0" level is below threshold ($I_0 < I_{th}$), stimulated emission is negligible until $N$ reaches $N_{th}$ and Eq. (1) is linear. For an RC-filtered step from $I_0$ to $I_1$ starting at $t = 0$ from the steady state $N(0) = I_0\tau_n/qV$,

$$N(t) = \frac{\tau_n}{qV}\left(I_0 + (I_1 - I_0)\left[1 - \frac{\tau_n e^{-t/\tau_n} - \tau_{RC} e^{-t/\tau_{RC}}}{\tau_n - \tau_{RC}}\right]\right), \tag{18}$$

and the carrier turn-on delay $t_d$ is the root of $N(t_d) = N_{th}$. For $\tau_{RC} \to 0$ this gives the classical result $t_d = \tau_n \ln[(I_1 - I_0)/(I_1 - I_{th})]$ [1,3]; a finite $\tau_{RC}$ always increases the delay.

## 5. Closed-form solutions for sinusoidal modulation

### 5.1 Power and chirp

For $I(t) = I_b + I_m \sin\Omega t$ with $\Omega = 2\pi f_m$ and $I_m$ small, Eq. (12) gives $p(t) = m_p \cos x$ with $x = \Omega t + \theta'$ and

$$m_p\, e^{i\theta} = \frac{a_{21} I_m}{qV\bar{S}\, Q(i\Omega)}, \qquad \theta' = \theta - \frac{\pi}{2}. \tag{19}$$

Equation (13) then gives a sinusoidal chirp that leads the power by the angle $\psi$:

$$\Delta\nu(t) = \nu_0 + \Delta\nu_m \cos(x + \psi), \tag{20}$$

$$\Delta\nu_m = \frac{\alpha}{4\pi} \zeta m_p \sqrt{\Omega^2 + \omega_c^2}, \qquad \psi = \arctan\frac{\Omega}{\omega_c}, \tag{21}$$

where $\nu_0 = (\alpha/4\pi)\left[\Gamma g_0\left(\bar{N} - N_0\right) - 1/\tau_p\right]$ is the adiabatic frequency offset of the bias point. The phase is $\phi(t) = 2\pi\nu_0 t + \beta_{FM}\sin(x + \psi)$, with frequency-modulation (FM) index $\beta_{FM} = \Delta\nu_m / f_m$ (not to be confused with the spontaneous-emission factor $\beta$). For $f_m \ll f_c$ the chirp is in phase with the power (adiabatic regime); for $f_m \gg f_c$ it is in quadrature (transient regime).

### 5.2 Line expansion of the Wigner distribution

The field is periodic apart from the carrier $e^{i2\pi\nu_0 t}$ and can be expanded as $E(t) = \sqrt{\bar{P}}\, e^{i2\pi\nu_0 t} \sum_n c_n\, e^{inx}$, with

$$c_n = \frac{1}{2\pi}\int_0^{2\pi} \sqrt{1 + m_p \cos x}\;\; e^{i\beta_{FM}\sin(x+\psi)}\, e^{-inx}\, dx. \tag{22}$$

To first order in $m_p$,

$$c_n \simeq e^{in\psi}\left(J_n(\beta_{FM}) + \frac{m_p}{4}\left[J_{n-1}(\beta_{FM}) e^{-i\psi} + J_{n+1}(\beta_{FM}) e^{i\psi}\right]\right), \tag{23}$$

where $J_n$ are Bessel functions; the asymmetry of the sidebands arising from the combined AM and FM is the classical result of [12,13]. Substituting the expansion into the definition of the pseudo-WVD (Section 7) with lag window $h(\tau)$ gives

$$C_h(t,f) = \bar{P} \sum_{n,k} c_n\, c_k^*\, e^{i(n-k)(\Omega t + \theta')}\, \hat{h}\left(f - \nu_0 - \frac{n+k}{2} f_m\right), \tag{24}$$

where $\hat{h}$ is the Fourier transform of $h$. For the WVD ($h \equiv 1$), $\hat{h}$ becomes a Dirac delta: the distribution is supported on lines at half-integer multiples of $f_m$; the lines with $n + k$ even carry the auto-terms ($n = k$) and cross-terms oscillating at multiples of $2f_m$, and the lines with $n + k$ odd are pure interference oscillating at odd multiples of $f_m$. Two regimes follow from Eq. (24). If $T_h \gg 1/f_m$, $\hat{h}$ is narrower than the line spacing $f_m/2$ and the distribution resolves the individual lines. If $T_h \ll 1/f_m$, the lines merge and, to leading order in $f_m T_h$, $C_h(t,f) \simeq P(t)\, \hat{h}[f - \Delta\nu(t)]$: the distribution is a band of width of order $1/T_h$ whose center follows the instantaneous frequency. The oscillation of this band is conspicuous only when the peak-to-peak excursion $2\Delta\nu_m$ is comparable to or larger than the band width; with $T_h = c/f_m$ and $c < 1$ this requires $\beta_{FM} \gtrsim 1/(2c)$, i.e., an FM index of order unity or larger. For $\beta_{FM} \ll 1$ the frequency modulation is still carried exactly by the first conditional moment (Section 7), but the oscillation of the band is a small fraction of its width.

## 6. Two-tone modulation, noise and dynamic range

Analog radio-over-fiber and microwave-photonic links are characterized by their intermodulation distortion and spurious-free dynamic range (SFDR) [33–35]. Intermodulation in directly modulated lasers originates in the nonlinear coupling of carriers and photons and is resonantly enhanced near $f_R$ [36,37]. In this section the single-tone analysis of Section 5 is extended to two tones by a Volterra expansion of Eqs. (1) and (2) [38,39], and the noise floor is obtained from a Langevin treatment of the same equations.

### 6.1 Volterra kernels of the rate equations

With $u = \delta N$, $v = \delta S$ and $h(S) = S/(1 + \varepsilon S)$, the only nonlinearity of Eqs. (1) and (2) is the stimulated term $g_0(N - N_0)h(S)$, which enters the carrier equation with coefficient $-1$ and the photon equation with coefficient $+\Gamma$, i.e., along $\mathbf{e} = (-1, \Gamma)^T$. Its second- and third-order terms around the bias point are

$$q_2 = g_0\left[h_1 uv + \frac{n}{2}h_2 v^2\right], \qquad q_3 = g_0\left[\frac{h_2}{2}uv^2 + \frac{n}{6}h_3 v^3\right], \tag{25}$$

where $n = \overline{N} - N_0$ and $h_1 = \left(1 + \varepsilon\overline{S}\right)^{-2}$, $h_2 = -2\varepsilon\left(1 + \varepsilon\overline{S}\right)^{-3}$ and $h_3 = 6\varepsilon^2\left(1 + \varepsilon\overline{S}\right)^{-4}$ are the derivatives of $h$ at $\overline{S}$. The linear part is the matrix of Eq. (8). We denote by $B_2$ and $T_3$ the symmetric bilinear and trilinear forms associated with $q_2$ and $q_3$, acting on vectors $\mathbf{x} = (x_u, x_v)$:

$$B_2(\mathbf{x}, \mathbf{y}) = g_0\left[\frac{h_1}{2}(x_u y_v + x_v y_u) + \frac{nh_2}{2}x_v y_v\right], \tag{26}$$

$$T_3(\mathbf{x}, \mathbf{y}, \mathbf{z}) = g_0\left[\frac{h_2}{6}(x_u y_v z_v + y_u x_v z_v + z_u x_v y_v) + \frac{nh_3}{6}x_v y_v z_v\right]. \tag{27}$$

With $\mathsf{M}(\omega) = (i\omega\mathsf{I} - \mathsf{A})^{-1}$, $\mathsf{A} = \left[a_{ij}\right]$ and $\mathbf{b} = (1/qV, 0)^T$, harmonic probing gives the first-, second- and third-order Volterra kernels of the state vector [39]:

$$\mathbf{H}_1(\omega_a) = \mathsf{M}(\omega_a)\,\mathbf{b}, \tag{28}$$

$$\mathbf{H}_2(\omega_a, \omega_b) = \mathsf{M}(\omega_a + \omega_b)\,\mathbf{e}\,B_2[\mathbf{H}_1(\omega_a), \mathbf{H}_1(\omega_b)], \tag{29}$$

$$\mathbf{H}_3(\omega_a, \omega_b, \omega_c) = \mathsf{M}(\omega_a + \omega_b + \omega_c)\,\mathbf{e}\,c_3(\omega_a, \omega_b, \omega_c), \tag{30}$$

$$c_3 = \frac{2}{3}\sum\nolimits_{\mathrm{cyc}} B_2\,[\mathbf{H}_1(\omega_a), \mathbf{H}_2(\omega_b, \omega_c)] + T_3[\mathbf{H}_1(\omega_a), \mathbf{H}_1(\omega_b), \mathbf{H}_1(\omega_c)], \tag{31}$$

where the sum runs over the three cyclic permutations of $(\omega_a, \omega_b, \omega_c)$. The kernels contain gain compression and spontaneous emission exactly, through $\mathsf{A}$ and the coefficients $h_k$. For the two-tone drive $\delta I(t) = I_m(\cos\omega_1 t + \cos\omega_2 t)$ (the phases of the tones do not affect the amplitudes below), the real amplitudes of the photon-density components at the fundamental and at the second- and third-order intermodulation frequencies are

$$\left|v_{f_1}\right| = I_m\left|H_{1,v}(\omega_1)\right|, \qquad \left|v_{f_1+f_2}\right| = I_m^2\left|H_{2,v}(\omega_1, \omega_2)\right|, \tag{32}$$

$$\left|v_{2f_1-f_2}\right| = \frac{3}{4}I_m^3\left|H_{3,v}(\omega_1, \omega_1, -\omega_2)\right|, \tag{33}$$

and the second harmonic is $\left|v_{2f_1}\right| = (I_m^2/2)\left|H_{2,v}(\omega_1, \omega_1)\right|$. The output-power amplitudes follow from $P = P_S S$. Because the chirp is linear in $u$ [Eq. (3)], the $u$ components of the same kernels, multiplied by $(\alpha/4\pi)\Gamma g_0$, give the intermodulation of the frequency modulation.

### 6.2 Relative intensity noise

The noise floor of the link is dominated by the relative intensity noise (RIN) of the laser, which we obtain from the Langevin form of the rate equations [2,40,41]. In terms of carrier and photon numbers, $N_c = NV$ and $N_p = SV/\Gamma$, the shot-noise Langevin forces, with correlations $\langle F_i(t)F_j(t')\rangle = D_{ij}\,\delta(t - t')$, have the diffusion coefficients

$$D_{pp} = \left(2n_{sp} - 1\right)GN_p + R_{sp} + \frac{N_p}{\tau_p}, \tag{34}$$

$$D_{nn} = (2n_{sp} - 1)GN_p + \frac{N_c}{\tau_n} + \frac{I}{q}, \tag{35}$$

$$D_{np} = -(2n_{sp} - 1)GN_p - R_{sp}, \tag{36}$$

where $G = \Gamma g_0(\overline{N} - N_0)/(1 + \varepsilon\overline{S})$ is the net modal gain rate, split into stimulated emission $n_{sp}G$ and absorption $(n_{sp} - 1)G$; $R_{sp} = \beta N_c/\tau_n$ is the spontaneous emission into the mode; and the pump is taken as shot-noise limited, which is pessimistic for a laser driven from a 50 Ω source at low frequencies. Transforming to densities with $\mathsf{T} = \mathrm{diag}(1/V, \Gamma/V)$, the single-sided RIN is

$$\mathrm{RIN}(f) = \frac{2}{\overline{S}^2}\left[\mathsf{M}(2\pi f)\,\mathsf{TDT}\,\mathsf{M}^{\dagger}(2\pi f)\right]_{vv}. \tag{37}$$

Equation (37) is the RIN of the intracavity photon number. As is common in link analyses, we use it for the output power and add the shot noise of the detected photocurrent separately in Eq. (39). This neglects the partition noise at the output coupler, which is accurate when the RIN exceeds the shot-noise-equivalent level $2h\nu/\overline{P}$ ($\approx -166$ dB/Hz at 10 mW) and becomes approximate at low frequencies and high bias, where the two are comparable. The phase is driven by the Langevin source $F_\phi$, uncorrelated with $F_N$ and $F_S$, with diffusion coefficient

$$D_{\phi\phi} = \frac{n_{sp}G}{2N_p}, \qquad \Delta\nu = (1 + \alpha^2)\frac{D_{\phi\phi}}{2\pi}, \tag{38}$$

where the second expression is the Lorentzian linewidth (full width at half maximum), including the $\alpha$-enhancement produced by the carrier fluctuations through Eq. (3) [7]. For the DFB-1550 laser, Eq. (38) gives 4.5, 1.5 and 0.9 MHz at 1.5, 2.5 and $3.5I_{th}$. A stochastic integration of Eqs. (1)–(3) with all three noise sources gives $1.54 \pm 0.02$ MHz at $2.5I_{th}$, i.e., within 4% of Eq. (38), from the variance of the phase increments over 2 ns. Phase noise does not affect the intensity-modulation link considered below, but it sets the coherence of the field and therefore the noise-induced spreading of its time–frequency distribution. The inversion factor $n_{sp}$ is an additional noise parameter (Table 3). As usual in rate-equation modeling, $\beta$ is treated as an effective parameter of the deterministic dynamics and is not tied to $n_{sp}$ by the Einstein relation.

### *6.3 Link model and spurious-free dynamic range*

We consider a back-to-back intensity-modulation/direct-detection link without dispersion (Table 3). The RF input power per tone delivered to a matched laser is $P_{in} = I_m^2 R_{in}/2$. The detected photocurrent is $i(t) = \mathcal{R}P(t)$, and the photodiode is treated as an unmatched current source delivering its full current into the load $R_L$, so that the RF power of a component of optical-power amplitude $\hat{P}$ is $P_{RF} = (\mathcal{R}\hat{P})^2 R_L/2$; with a resistively matched photodiode [33], gain, intercepts and the shot and RIN noise terms are 6 dB lower. The output noise power spectral density is

$$\mathcal{N}(f) = k_B T + 2qI_{dc}R_L + \mathrm{RIN}(f)\,I_{dc}^2 R_L, \qquad I_{dc} = \mathcal{R}\overline{P}. \tag{39}$$

Because Eq. (33) gives the exact small-signal scaling of each product, the output intercept points follow directly from the kernels, $\mathrm{OIP}_2 = 2P_{f_1} - P_{f_1+f_2}$ and $\mathrm{OIP}_3 = (3P_{f_1} - P_{2f_1-f_2})/2$ (all in dBm and independent of $I_m$ in the small-signal limit). The SFDRs in a 1 Hz noise bandwidth are then [33]

$$\mathrm{SFDR}_2 = \frac{1}{2}[\mathrm{OIP}_2 - \mathcal{N}(f_1 + f_2)], \tag{40}$$

$$\mathrm{SFDR}_3 = \frac{2}{3}[\mathrm{OIP}_3 - \mathcal{N}(2f_1 - f_2)], \tag{41}$$

in dB·Hz$^{1/2}$ and dB·Hz$^{2/3}$, respectively, with $\mathcal{N}$ expressed in dBm/Hz and evaluated at the frequency of the corresponding distortion product.

**Table 3. Link and noise parameters used for the two-tone analysis.**

| Parameter | Symbol | Value |
|---|---|---|
| photodiode responsivity | $\mathcal{R}$ | 0.8 A/W (0.6 A/W for the 850 nm link) |
| photodiode termination | – | unmatched, full photocurrent into $R_L$ |
| load and laser source resistance (matched laser input) | $R_L$, $R_{in}$ | 50 Ω |
| optical loss between laser and detector | – | 0 dB (back-to-back) |
| temperature | $T$ | 290 K |
| population-inversion factor | $n_{sp}$ | 2 |
| tone spacing | $f_2 - f_1$ | 50 MHz |
| laser bias | $I_b$ | 1.5, 2.5, 3.5 $I_{th}$ |

## 7. Time–frequency framework

### *7.1 Cohen's class*

For the envelope $E(t)$ we define the instantaneous autocorrelation $K(t,\tau) = E(t+\tau/2)E^*(t-\tau/2)$ and the symmetric ambiguity function $\mathcal{A}(\nu,\tau) = \int K(t,\tau)e^{-i2\pi\nu t}dt$. Cohen's class of quadratic time–frequency distributions is [16,17]

$$C(t,f) = \iint \Phi(\nu,\tau)\,\mathcal{A}(\nu,\tau)\,e^{i2\pi(\nu t - f\tau)}\,d\nu\,d\tau, \tag{42}$$

where the kernel $\Phi$ selects the member. The WVD has $\Phi = 1$; the pseudo-WVD (PWVD) has $\Phi = h(\tau)$, a lag window that smooths along frequency; the smoothed pseudo-WVD (SPWVD) has $\Phi = \hat{g}(\nu)h(\tau)$, where $\hat{g}$ is the Fourier transform of a unit-area time-smoothing window $g(t)$; the Choi–Williams distribution (CWD) has $\Phi = \exp[-(2\pi\nu\tau)^2/\sigma]$ [42]; and the spectrogram, $\left|\int E(u)w(u-t)e^{-i2\pi f u}du\right|^2$, has a kernel equal to the ambiguity function of the analysis window $w$ (for a real, even window) [18]. The kernels used in this work and their properties are summarized in Table 4. We use Gaussian windows throughout and specify them by their full width at half maximum (FWHM): $T_h$ is the FWHM of $h(\tau)$, normalized to $h(0) = 1$; $T_g$ that of $g(t)$; and $T_w$ the FWHM of $|w|^2$ for the spectrogram.

**Table 4. Cohen-class distributions used in this work. "IF" indicates that the first conditional moment equals $\phi'/2\pi$ (Eq. (44)); "cross-terms" indicates the direction in the ambiguity plane in which interference is attenuated.**

| Distribution | Kernel $\Phi(\nu,\tau)$ | Time marginal | Freq. marginal | IF | Cross-terms attenuated |
|---|---|---|---|---|---|
| WVD | 1 | yes | yes | yes | none |
| PWVD | $h(\tau)$ | yes | no | yes | large lags $\lvert\tau\rvert \gtrsim T_h$ |
| SPWVD | $\hat{g}(\nu)h(\tau)$ | no | no | no | large $\lvert\tau\rvert$ and $\lvert\nu\rvert$ |
| CWD | $e^{-(2\pi\nu\tau)^2/\sigma}$ | yes | yes | yes | away from both axes |
| Spectrogram | ambiguity function of $w$ | no | no | no | all (non-negative) |

With an additional lag window (Figs. 5(c) and 14), the CWD keeps the time marginal and the IF property but not the exact frequency marginal.

### *7.2 Marginals and the first conditional moment*

The following properties are well known [17,20]; we recall them because they fix which distributions can be used quantitatively. Integrating Eq. (42) over $f$ produces $\delta(\tau)$, and integrating over $t$ produces $\delta(\nu)$. Hence

$$\int C\,df = P(t) \Leftrightarrow \Phi(\nu,0) = 1, \qquad \int C\,dt = \left|\tilde{E}(f)\right|^2 \Leftrightarrow \Phi(0,\tau) = 1, \tag{43}$$

where $\tilde{E}$ is the Fourier transform of the field. The time marginal is thus the optical power and the frequency marginal is the optical energy spectrum. For the first moment, $\int f e^{-i2\pi f\tau} df = (i/2\pi)\delta'(\tau)$, so that $\int fC\,df = -(i/2\pi)\int \partial_\tau[\Phi\mathcal{A}]_{\tau=0} e^{i2\pi\nu t} d\nu$. If $\Phi(\nu,0) = 1$ and $\partial_\tau \Phi(\nu,0) = 0$, only $\partial_\tau K(t,0) = 1/2\,(E'E^* - EE'^*) = iP\phi'$ survives, and

$$\langle f \rangle_t \equiv \frac{\int fC(t,f)\,df}{\int C(t,f)\,df} = \frac{\phi'(t)}{2\pi} = \Delta\nu(t). \tag{44}$$

The laser chirp is therefore the exact centroid, at each instant, of the WVD, the PWVD with a symmetric lag window, and the CWD. The SPWVD and the spectrogram violate $\Phi(\nu,0) = 1$ and return a power-weighted, time-smoothed chirp instead [17,20].

### *7.3 Instantaneous bandwidth and the geometry of chirp*

Writing $E = A\,e^{i\phi}$ with $A = \sqrt{P}$, the rms bandwidth $B$ of the optical spectrum about its mean frequency $\overline{\nu}$ satisfies Cohen's bandwidth equation [17]

$$B^2 = \frac{1}{4\pi^2\mathcal{E}}\int A'^2\,dt + \frac{1}{\mathcal{E}}\int [\Delta\nu(t) - \overline{\nu}]^2 P(t)\,dt, \qquad \mathcal{E} = \int P\,dt, \tag{45}$$

which identifies $B_s(t) = |A'/A|/2\pi = |d\ln P/dt|/4\pi$ as the natural instantaneous bandwidth [17]. Note that $B_s$ is defined through Eq. (45) and does not coincide with the conditional spread of a particular distribution: for the WVD, for instance, the conditional variance is $1/2\,[(A'/A)^2 - A''/A]/4\pi^2$ and can be negative, and for the PWVD it depends on $T_h$. Since $d\ln P/dt = 2A'/A$, the transient term of Eq. (6) can be written

$$\Delta\nu_{tr}(t) = \alpha\left(1 + \tau_p\kappa P\right)\frac{A'(t)}{2\pi A(t)}, \qquad |\Delta\nu_{tr}| = \alpha(1 + \varepsilon S)B_s(t). \tag{46}$$

The transient chirp is therefore $\alpha(1 + \varepsilon S)$ times the instantaneous bandwidth, which reduces to $\alpha B_s$ for $\varepsilon S \ll 1$. Equation (46) is an algebraic restatement of Eq. (6), but it gives the $\alpha$-factor a geometric meaning in terms of Cohen's quantities: it is the ratio between the displacement of the centroid produced by an amplitude transition and the instantaneous bandwidth that the same transition generates.

**rms-bandwidth budget.** Let the pulse have finite energy and start and end at the same power level (e.g., a pulse on a dark background; a transition between steady states requires a windowed record), and let $\beta \to 0$. Substituting $\Delta\nu = \Delta\nu_{tr} + \Delta\nu_{ad}$ with $\Delta\nu_{tr}$ from Eq. (46) and $\Delta\nu_{ad} = (\alpha/4\pi)\kappa A^2$ into Eq. (45), the cross term is proportional to $\int \left(1 + \tau_p\kappa A^2\right)A'A(\Delta\nu_{ad} - \overline{\nu})\,dt$. Its integrand is the time derivative of a polynomial in $A^2$, so the integral vanishes between equal end points. For the same reason the transient term does not contribute to the mean frequency, so that $\overline{\nu} = \langle \Delta\nu_{ad} \rangle$, the power-weighted mean of the adiabatic chirp. Hence

$$B^2 = B_{TL}^2 + B_{tr}^2 + \sigma_{ad}^2, \tag{47}$$

$$B_{TL}^2 = \frac{\int A'^2 dt}{4\pi^2\mathcal{E}}, \qquad B_{tr}^2 = \frac{\alpha^2\int (1+\varepsilon S)^2 A'^2 dt}{4\pi^2\mathcal{E}}, \qquad \sigma_{ad}^2 = \frac{\int (\Delta\nu_{ad} - \overline{\nu})^2 P\,dt}{\mathcal{E}}, \tag{48}$$

where $B_{TL}$ is the rms bandwidth of the transform-limited pulse with the same intensity profile. For $\varepsilon S \ll 1$, $B_{tr}^2 \to \alpha^2 B_{TL}^2$ and $B^2 = (1 + \alpha^2)B_{TL}^2 + \sigma_{ad}^2$. The broadening factor $\sqrt{1 + C^2}$ of Gaussian pulses with linear chirp parameter $C$ is thus a special case: for any pulse shape, transient chirp broadens the rms spectrum by $\sqrt{1 + \alpha^2}$ (in the weak-compression limit), and adiabatic chirp adds its power-weighted variance in quadrature.

### *7.4 Parameter extraction from the distribution*

All the regressors in Eq. (6) are functionals of the time marginal $P(t)$, and the left-hand side is the first conditional moment. For any kernel satisfying the conditions of Eq. (44), the model

$$\langle f\rangle_t = a_1 \frac{d\ln P}{dt} + a_3\, P \frac{d\ln P}{dt} + a_2 P + a_0 \tag{49}$$

is exact for $\beta \to 0$, with $a_1 = \alpha/4\pi$, $a_2 = \alpha\kappa/4\pi$ and $a_3 = \alpha\tau_p\kappa/4\pi$. A linear least-squares fit over the samples where the power exceeds 5% of its peak therefore yields $\alpha = 4\pi a_1$, the adiabatic coefficient $\alpha\kappa/4\pi = a_2$ and the photon lifetime $\tau_p = a_3/a_2$ from the distribution alone. The regressor columns differ by more than ten orders of magnitude in SI units and must be normalized before solving. We refer to Eq. (49) as the four-term estimator and to the Koch–Linke form ($a_3 = 0$) as the three-term estimator. Because $\langle f\rangle_t = \phi'/2\pi$ exactly for these kernels, the regression is mathematically equivalent to a Koch–Linke analysis of the time-resolved chirp, as used with chirp measurements [29,30,43]; the time–frequency formulation contributes the kernel conditions that determine when moment-based chirp estimates are unbiased, which matters for spectrographic data. The identification $\tau_p = a_3/a_2$ relies on the index-neutrality of gain compression assumed in Eq. (3); a compression-induced index change would add a term proportional to $P$ in the chirp and change the meaning of $a_3/a_2$.

### *7.5 Discrete implementation*

The field is sampled with interval $\Delta t$. With $K[j,m] = E[j+m]E^*[j-m]$ the lag is $\tau_m = 2m\Delta t$, and the PWVD is computed as

$$C[j,l] = 2\Delta t \sum_{m=-M}^{M} h\,(2m\Delta t) K[j,m]\, e^{-i2\pi lm/L}, \tag{50}$$

at the frequencies $f_l = l/(2L\Delta t)$, where $L$ is the FFT length. The discrete WVD is periodic in frequency with period $1/(2\Delta t)$, so a complex envelope is represented without aliasing only if its spectrum lies within $|f| < 1/(4\Delta t)$ [44]. General kernels are applied in the ambiguity domain by an FFT along $j$ with zero padding to twice the record length, multiplication by $\Phi$, and an inverse FFT. A practical subtlety concerns the first moment on the periodic discrete frequency axis: $\sum_l f_l\, C[j,l]$ weights the lag samples by a factor that decays only as $1/m$, so that, unlike its continuous counterpart, the discrete WVD centroid is not local. In low-power intervals it can be biased by cross-terms generated by distant high-power events. A lag window restores locality, which is why the PWVD is the preferred estimator of the chirp.

## 8. Numerical method

The rate equations (1)–(3) are integrated with the LSODA method [45] as implemented in SciPy [46], using a relative tolerance of $10^{-9}$, per-component absolute tolerances scaled to the steady state, a maximum step of 2 ps and scaled state variables. Every record is preceded by 3 ns (10 ns for sub-threshold biasing) at the initial current so that it starts from the steady state. The output is sampled at $\Delta t = 1$ ps, which makes the discrete WVD alias-free for $|f| < 250$ GHz, far beyond the chirp excursions encountered. Small-signal amplitudes and phases are obtained by synchronous (lock-in) demodulation over an integer number of modulation periods after the turn-on transient has decayed. The parameter regression uses the samples where the power exceeds 5% of its peak and excludes the first and last 0.15 ns of each record (the first 2 ns for sinusoidal records, to discard the turn-on transient). To assess robustness, the field is passed through a Gaussian receiver filter of 3-dB bandwidth $B_{rx}$, and complex white Gaussian noise filtered by the same response is added at a signal-to-noise ratio (SNR) defined as the ratio of the mean signal power to the noise power within $B_{rx}$; 20 noise realizations are used per SNR value. The drive conditions used in the figures are summarized in Table 5.

**Table 5. Drive conditions used in the simulations (DFB-1550 unless stated otherwise).**

| Figure | Waveform | Levels | Timing |
|---|---|---|---|
| 2 | sinusoid, small signal | $I_b = 1.5{,}2.5{,}3.5 I_{th}$; $I_m = 0.01 I_{th}$ | $f_m$ = 0.5–14 GHz |

| Figure | Waveform | Levels | Timing |
|---|---|---|---|
| 3 | RC-filtered pulse, small signal | $I_b = 2.5I_{th}$; $\delta I = 0.01I_{th}$ | $T = 200$ ps; $\tau_{RC} =$ 5,20,40 ps |
| 4, 5 | RC-filtered pulse, large signal | $I_0 = 1.2I_{th}$, $I_1 = 3.5I_{th}$ | $T = 100$ ps; $\tau_{RC} = 15$ ps |
| 6 | RC-filtered pulse, large signal | $I_0 = 1.2I_{th}$, $I_1 = 3.5I_{th}$ | $T = 100$ ps; $\tau_{RC} =$ 2,15,40 ps |
| 7, 14 | RC-filtered NRZ, 16 bits | $I_0 = 1.2I_{th}$, $I_1 = 3.5I_{th}$ | 10 Gb/s; $\tau_{RC} = 15$ ps |
| 8 | sinusoid | $I_b = 2.5I_{th}$; $I_m = 0.1,0.6I_{th}$ | $f_m = 1$ GHz, $0.8f_R =$ 5.17 GHz |
| 9 | sinusoid | $I_b = 2.5I_{th}$; $I_m = 0.1I_{th}$ | $f_m = 5.17$ GHz |
| 10 | sinusoid, large signal | $I_b = 2.5I_{th}$; $I_m = 1.3I_{th}$ | $f_m = 1$, 5.17 GHz |
| 11 | two tones | $I_b = 2.5I_{th}$; $I_m = 0.5I_{th}$ per tone | $f_1 = 2$, $f_2 = 2.25$ GHz |
| 12, 13 | two tones | $I_b = 1.5$–$3.5I_{th}$; kernels for $P_{in}$ from −60 to 5 dBm, simulations from −30 to 0 dBm per tone | $f_1 = 0.2$–10 GHz, $f_2 = f_1 + 50$ MHz |
| 15 | two RC-filtered pulses | $I_0 = 1.2I_{th}$, $I_1 = 3.5I_{th}$ | 600 ps apart; $\tau_{RC} = 15$ ps |
| 16 | RC-filtered NRZ, 8 bits, three lasers | $I_0 = 1.5I_{th}$, $I_1 = 5I_{th}$ | 10 Gb/s; $\tau_{RC} = 15$ ps |

## 9. Results

### 9.1 Validation of the small-signal solutions

Figure 2 compares the closed-form IM response, Eq. (12), and CPR, Eq. (13), with values obtained by lock-in demodulation of the numerically integrated rate equations at three bias currents. Across 36 frequency–bias combinations, the maximum deviation is 0.007 dB in $|H|$, 0.075% in $|\mathrm{CPR}|$ and 0.04° in its phase, i.e., at the level of the numerical tolerance. The CPR magnitude shows the expected transition from an adiabatic floor, 0.99, 2.89 and 4.81 GHz per unit relative modulation at the three biases, which corresponds to 0.280–0.286 GHz/mW (Table 2), to a transient slope of approximately $(\alpha/2)f$. The corner frequency increases with bias from 0.65 to 3.08 GHz because $\omega_c \approx \kappa\overline{P}$.

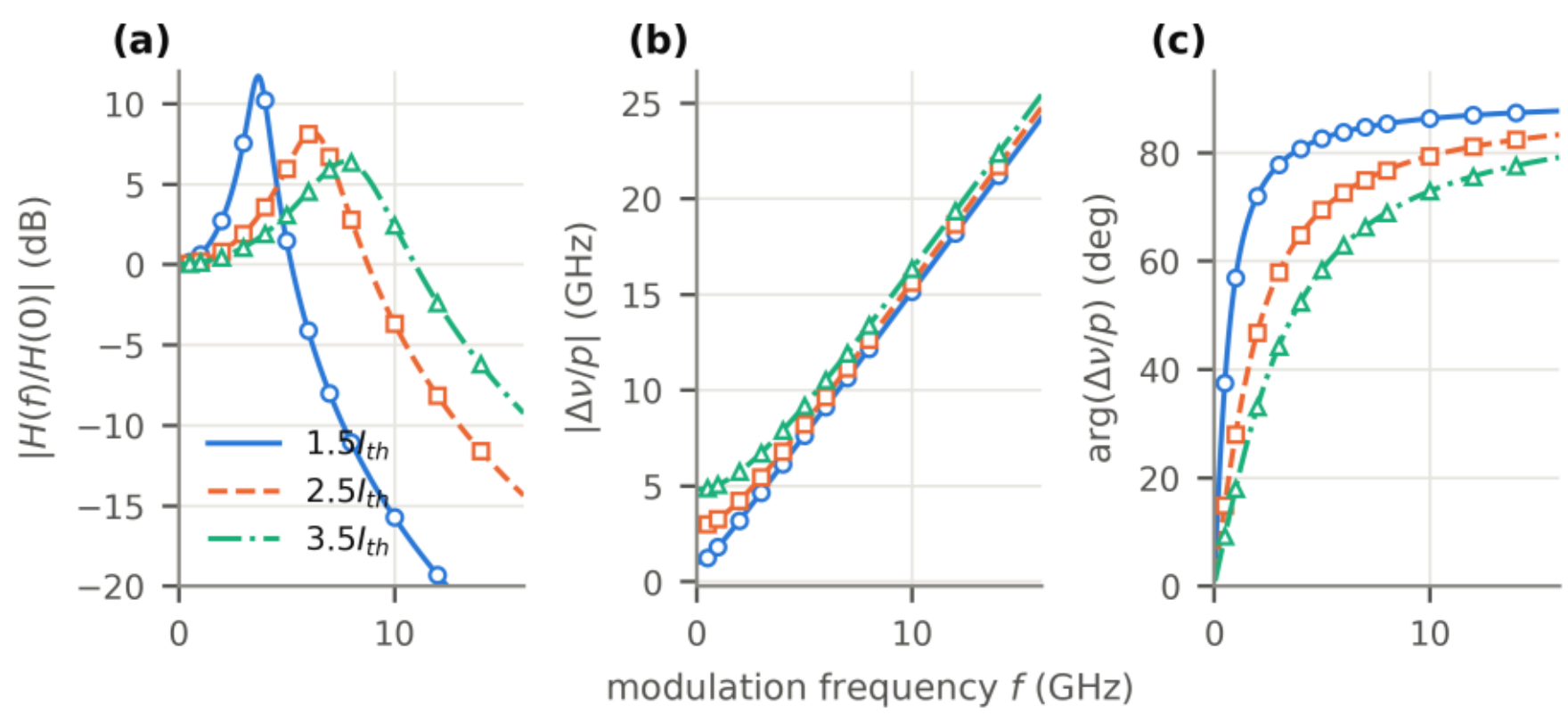

**Fig. 2.** Small-signal response of the DFB-1550 laser at $1.5I_{th}$ (solid, circles), $2.5I_{th}$ (dashed, squares) and $3.5I_{th}$ (dash-dotted, triangles). (a) Normalized IM response $|H(f)|$; (b) magnitude and (c) phase of the chirp-to-power ratio $\delta\nu/p$. Lines: closed form, Eqs. (12) and (13); markers: numerical integration with lock-in demodulation.

Figure 3 validates the RC-pulse solution, Eqs. (15)–(17), for a 200 ps pulse of amplitude $0.01I_{th}$ on a $2.5I_{th}$ bias and three RC constants. The maximum relative errors are 0.23–0.38% in $p(t)$ and 0.40–0.58% in $\delta\nu(t)$. The error scales linearly with the pulse amplitude (1.57%, 0.32% and 0.065% for $\delta I = 0.05$, 0.01 and $0.002I_{th}$ at $\tau_{RC} = 20$ ps), confirming that the residual is the leading nonlinear correction and not an error of the linear solution. Increasing $\tau_{RC}$ from 5 to 40 ps reduces the peak chirp from 65 to 44 MHz and the relaxation overshoot, as predicted by the additional real pole in Eq. (15). The sub-threshold delay, Eq. (18), agrees with the numerical carrier trajectories within the 1 ps output grid for the cases listed in Table 6.

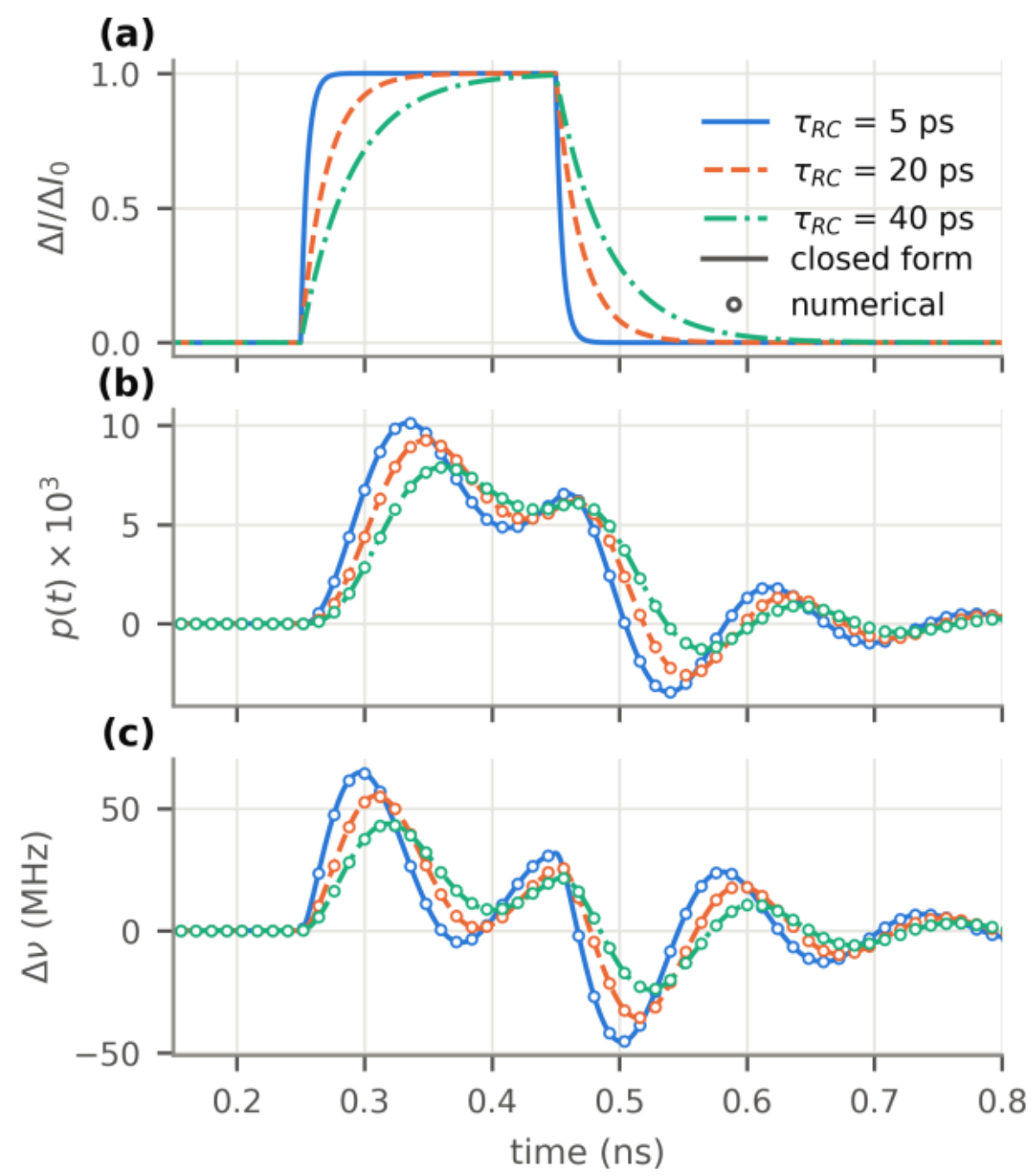

**Fig. 3.** Small-signal response to an RC-filtered rectangular current pulse ($T = 200$ ps, $\delta I = 0.01I_{th}$, $I_b = 2.5I_{th}$) for $\tau_{RC} = 5$ ps (solid), 20 ps (dashed) and 40 ps (dash-dotted). (a) Normalized current; (b) relative power modulation $p(t)$; (c) chirp $\delta\nu(t)$. Lines: closed form, Eqs. (15)–(17); circles: numerical integration.

**Table 6. Summary of the comparison between closed-form and numerical results.**

| Quantity | Closed form | Conditions | Deviation |
|---|---|---|---|
| IM response $\|H\|$ | Eq. (12) | 36 points, Fig. 2 | ≤ 0.007 dB |
| CPR magnitude / phase | Eq. (13) | 36 points, Fig. 2 | ≤ 0.075% / ≤ 0.04° |
| RC pulse, $p(t)$ / $\delta\nu(t)$ | Eqs. (15)–(17) | $\delta I = 0.01I_{th}$, Fig. 3 | ≤ 0.38% / ≤ 0.58% |
| Turn-on delay, $I_0 = 0.5I_{th}$, $\tau_{RC} = 20$ ps | Eq. (18) | $I_1 = 3I_{th}$ | 243.6 vs 244 ps |
| Turn-on delay, $I_0 = 0.8I_{th}$, $\tau_{RC} = 5$ / 30 ps | Eq. (18) | $I_1 = 3I_{th}$ | 100.4 / 125.2 vs 101 / 126 ps |
| Turn-on delay, $I_0 = 0.9I_{th}$, $\tau_{RC} = 15$ ps | Eq. (18) | $I_1 = 3I_{th}$ | 63.7 vs 64 ps |
| Sine, chirp amplitude | Eq. (20) | $I_m = 0.1I_{th}$, $f_m = 1$ / 5.17 GHz | 0.12% / 0.29% |

| Quantity | Closed form | Conditions | Deviation |
|---|---|---|---|
| Two-tone products (Volterra) | Eqs. (28)–(33) | $I_m = 0.02I_{th}$ / $P_{in} \le -10$ dBm | ≤ 0.11% / ≤ 0.11 dB |
| RIN, 1–11.5 GHz | Eq. (37) | Langevin simulation, $2.5I_{th}$ | ≤ 0.7 dB |
| Linewidth | Eq. (38) | Langevin simulation, $2.5I_{th}$ | 1.48 vs 1.54 ± 0.02 MHz |
| Sideband powers, $\lvert n\rvert \le 1$ | Eq. (22) | $I_m = 0.1I_{th}$, $f_m = 5.17$ GHz | ≤ 0.05 dB |
| PWVD, $T_h = 1$ ns | Eq. (24) | as above, Fig. 9 | 0.6% of peak |
| Time marginal (WVD, PWVD, CWD) | Eq. (43) | Fig. 5 | $4 \times 10^{-16}$ |
| rms bandwidth, Cohen identity | Eq. (45) | Fig. 14(d) | ≤ 0.05% |
| rms-bandwidth budget, $\beta = 10^{-4}$ / $10^{-6}$ | Eq. (47) | Fig. 14(d) | −0.5 to −1.9% / ≤ 0.08% |

### *9.2 Large-signal RC-filtered pulses*

Figure 4 shows the response of the reference laser to a single 100 ps RC-filtered pulse ($\tau_{RC} =$ 15 ps) from $1.2I_{th}$ to $3.5I_{th}$. The carrier density overshoots to $1.078N_{th}$ during the turn-on, producing a 35 mW optical spike, and undershoots to $0.961N_{th}$ after the trailing edge, which switches the laser almost off for about 0.4 ns. The chirp, computed with Eq. (5), reaches +20.9 GHz on the leading edge and −10.5 GHz on the trailing edge. The decomposition shows that the leading and trailing excursions are transient, whereas the adiabatic term, peaking at 9.8 GHz, follows the power spike; the spontaneous term appears only during the near-extinction interval.

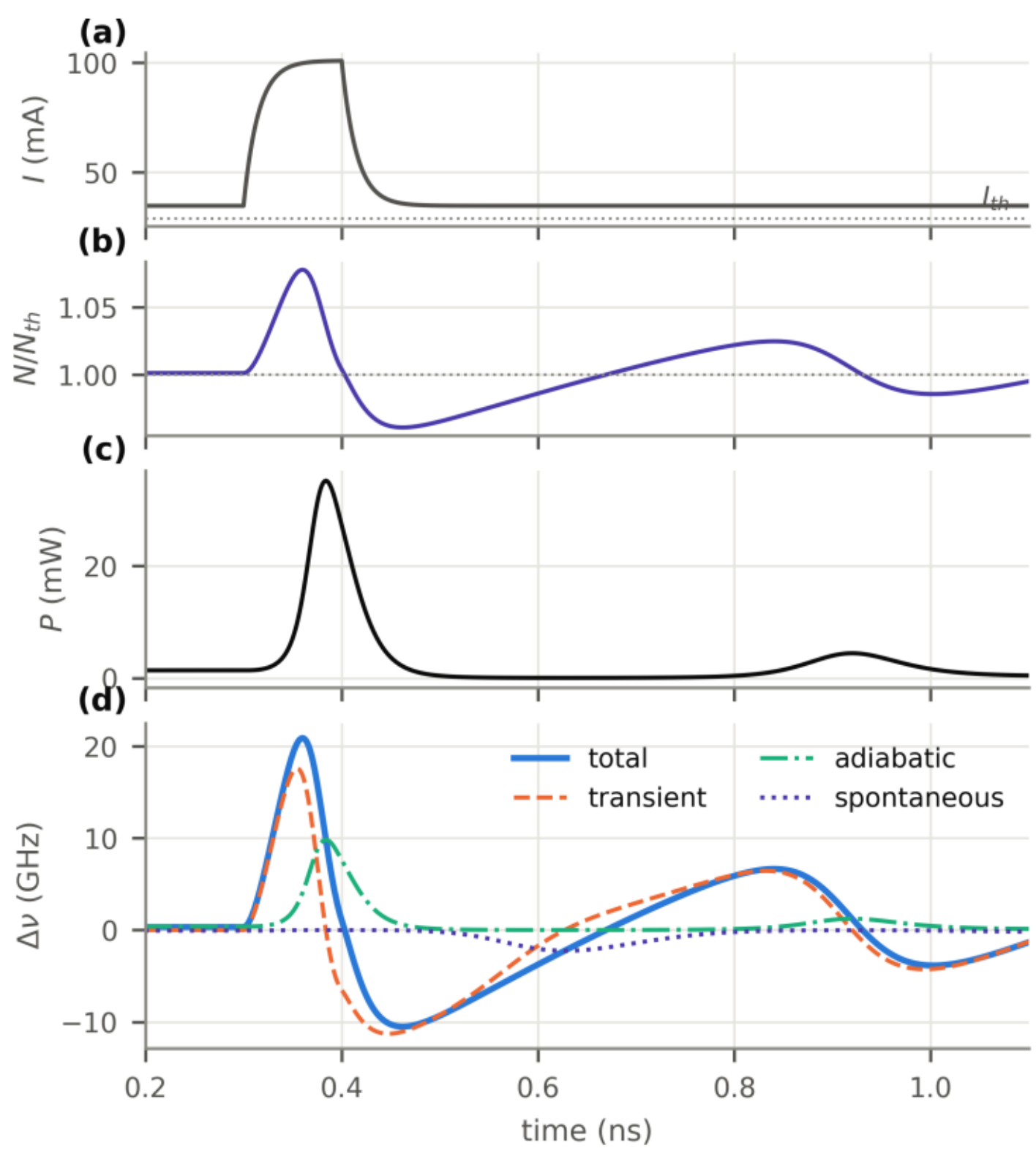

**Fig. 4.** Large-signal response of the DFB-1550 laser to an RC-filtered current pulse ($T = 100$ ps, $\tau_{RC} = 15$ ps, $1.2 \rightarrow 3.5I_{th}$). (a) Current; (b) normalized carrier density; (c) output power; (d)

chirp (thick solid) and its transient (dashed), adiabatic (dash-dotted) and spontaneous (dotted) components from Eq. (5).

The same pulse is represented in Fig. 5 by five members of Cohen's class. The WVD, Fig. 5(a), satisfies both marginals exactly: its time marginal reproduces $P(t)$ to $4 \times 10^{-16}$ (relative) and its frequency marginal coincides with the FFT energy spectrum to $4 \times 10^{-6}$, Fig. 5(f). However, it is dominated by interference between the pulse and the continuous-wave background before and after it, which appears as oscillating positive and negative regions [47]. A 40 ps lag window, Fig. 5(b), removes the long-lag interference and leaves a single ridge that follows the chirp from the blue-shifted leading edge to the red-shifted tail, while retaining the exact time marginal and centroid. The CWD, Fig. 5(c), which here includes a 200 ps lag window, retains the time marginal and the centroid but leaves interference located near the axes of the ambiguity plane (see Section 9.7). The SPWVD and the spectrogram, Figs. 5(d) and (e), are essentially non-negative but blur the chirp in time; their ridges are biased towards the chirp at the high-power instants, which dominate the smoothing.

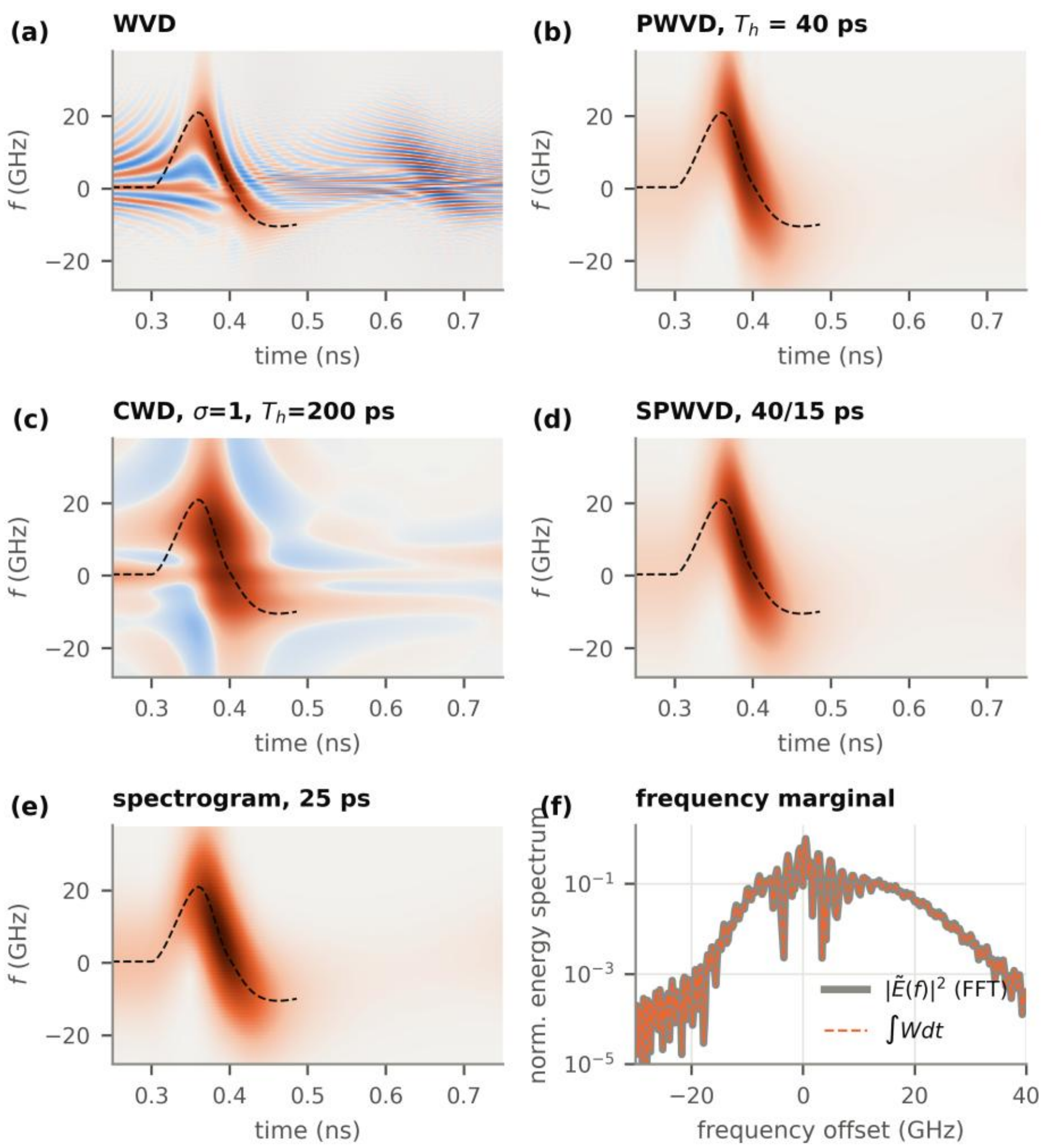

**Fig. 5.** The large-signal pulse of Fig. 4 represented by (a) the WVD; (b) the PWVD with $T_h =$ 40 ps; (c) the CWD with $\sigma = 1$ and a 200 ps lag window; (d) the SPWVD with $T_h = 40$ ps and $T_g = 15$ ps; (e) the spectrogram with $T_w = 25$ ps. Dashed lines: chirp from Eq. (3), shown where $P$ exceeds 2% of its peak. Color scale: $\mathrm{sign}(C)|C/C_{max}|^{1/2}$, warm colors positive and blue negative. (f) Frequency marginal of the WVD (thin dashed) and FFT energy spectrum (thick gray).

The effect of the driver bandwidth is shown in Fig. 6. As $\tau_{RC}$ increases from 2 to 40 ps, the peak power falls from 37.1 to 29.2 mW, the peak-to-peak chirp from 34.6 to 24.3 GHz and the power-weighted rms transient chirp from 8.3 to 6.4 GHz. In the PWVD, the slower current edge appears as a less steeply inclined ridge and a narrower frequency spread at the leading edge. The RC constant thus acts as a direct control of the slope of the time–frequency trajectory.

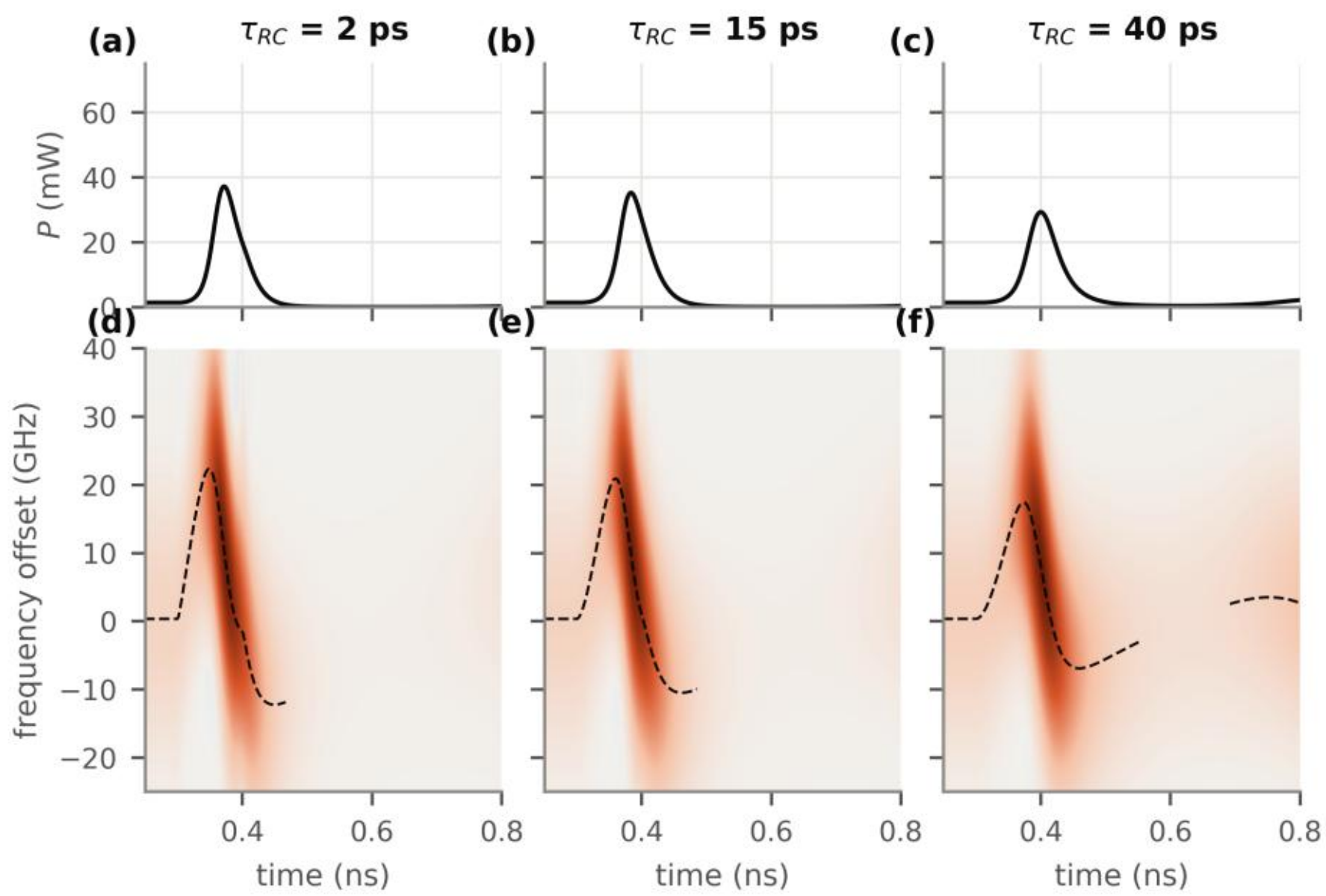

**Fig. 6.** Influence of the RC constant on a large-signal pulse ($T = 100$ ps, 1.2 → $3.5I_{th}$). (a)–(c) Output power and (d)–(f) PWVD ($T_h = 40$ ps) with the chirp (dashed) for $\tau_{RC} = 2$, 15 and 40 ps.

### *9.3 Mechanism atlas for RC-filtered NRZ modulation*

Figure 7 applies the PWVD to a 16-bit RC-filtered NRZ sequence at 10 Gb/s, switching the physical mechanisms on and off. With $\alpha = 0$, Fig. 7(b), the chirp vanishes identically and the distribution is symmetric about zero frequency; its spread in frequency is purely the instantaneous bandwidth of the amplitude transitions. With $\alpha = 3$ and negligible compression ($\varepsilon = 10^{-24}$ m³), Fig. 7(c), the relaxation oscillations are weakly damped, the power reaches 110 mW, and each spike produces a steep transient stroke spanning +41 to −33 GHz. At the nominal compression, Fig. 7(d), the strokes persist on isolated "1" bits, whereas consecutive "1" bits settle towards an adiabatic plateau. At strong compression ($\varepsilon = 5 \times 10^{-23}$ m³), Fig. 7(e), the relaxation oscillations are strongly damped, the peak power drops to 26 mW, and the "1" level is shifted by about 14.5 GHz with respect to the "0" level. The time–frequency picture thus separates at a glance the three regimes that determine the dispersion tolerance of a DML: symmetric (chirp-free), transient-dominated and adiabatic-dominated.

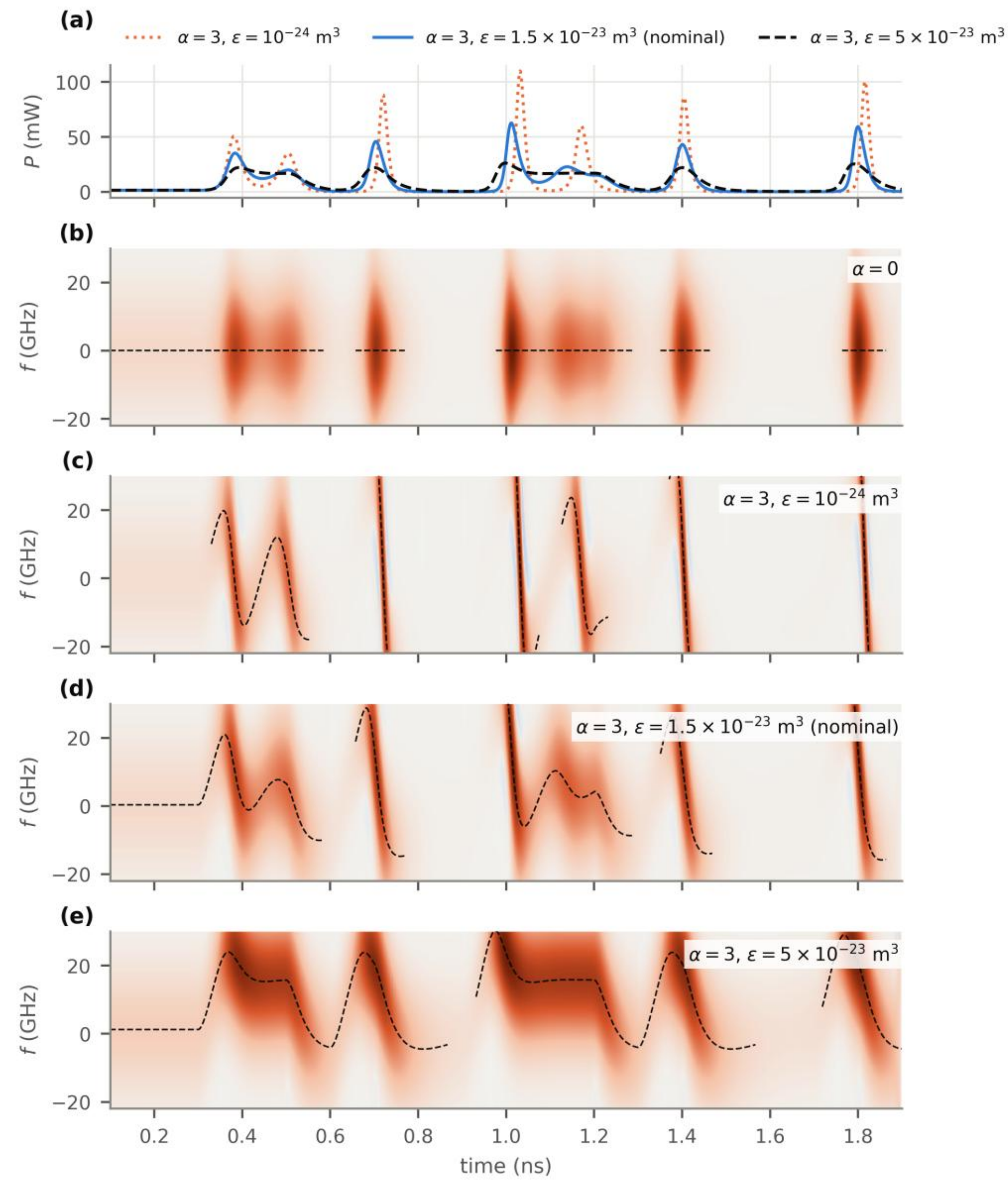

**Fig. 7.** Mechanism atlas for 10 Gb/s RC-filtered NRZ modulation ($\tau_{RC} = 15$ ps, $1.2 \rightarrow 3.5 I_{th}$). (a) Output power for $\varepsilon = 10^{-24}$ m³ (dotted), $1.5 \times 10^{-23}$ m³ (solid; identical for $\alpha = 0$) and $5 \times 10^{-23}$ m³ (dashed). (b)–(e) PWVD ($T_h = 40$ ps) with the chirp (dashed) for the parameter sets indicated in each panel.

### *9.4 Sinusoidal modulation*

Figure 8 shows the PWVD for sinusoidal modulation at $f_m = 1$ GHz (below both $f_c = 1.88$ GHz and $f_R$, i.e., in the adiabatic-dominated regime) and at $f_m = 0.8 f_R = 5.17$ GHz (transient-dominated), with $I_m = 0.1 I_{th}$ and $0.6 I_{th}$. The closed-form instantaneous frequency, Eq. (20), is superimposed on the numerical chirp. For $I_m = 0.1 I_{th}$ the FM index is $\beta_{FM} = 0.22$ and $0.23$, and the rms deviation from the numerical chirp is 2.7 and 32 MHz (1.2% and 2.7% of the chirp amplitude). Consistently with Section 5.2, the band oscillation (±0.22 and ±1.17 GHz) is small compared with the band width set by the lag windows (of order 6 and 15 GHz), although the centroid tracks the chirp. For $I_m = 0.6 I_{th}$ ($\beta_{FM} \approx 1.35$) the oscillation becomes comparable to the band width and a sinusoidal ridge is clearly visible. At 1 GHz it remains well described by the small-signal theory (8.6% rms deviation). At 5.17 GHz the relative power modulation reaches $m_p = 0.83$ and the numerical chirp amplitude, 7.9 GHz, exceeds the linear prediction of 7.0 GHz, which is the onset of large-signal behavior. The phase lead of the chirp over the power increases from $\psi = 28°$ at 1 GHz to 70° at 5.17 GHz, in agreement with Eq. (21).

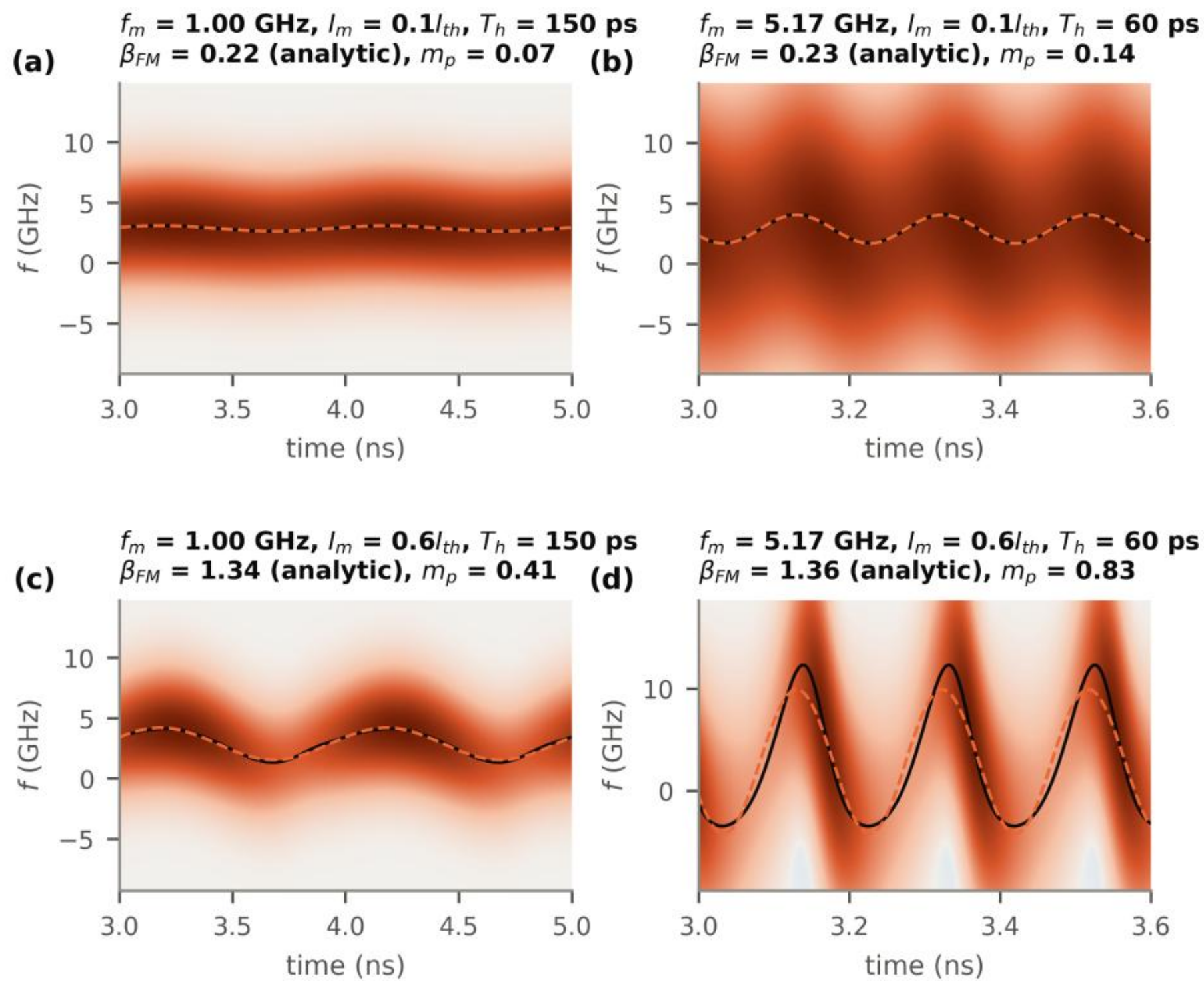

**Fig. 8.** PWVD for sinusoidal modulation on a $2.5I_{th}$ bias. (a) $f_m = 1$ GHz ($T_h = 150$ ps) and (b) $f_m = 5.17$ GHz ($T_h = 60$ ps) with $I_m = 0.1I_{th}$; (c), (d) the same with $I_m = 0.6I_{th}$. Solid black: numerical chirp; dashed orange: closed form, Eq. (20). The FM index $\beta_{FM}$ and power modulation depth $m_p$ from Eqs. (19)–(21) are indicated.

The line structure predicted by Eqs. (22) and (24) is verified in Fig. 9 for $f_m = 5.17$ GHz and $I_m = 0.1I_{th}$. The powers of the carrier and first sidebands obtained from the numerical field agree with $|c_n|^2$ to within 0.05 dB, including the asymmetry between the $n = -1$ (−19.5 dB) and $n = +1$ (−17.8 dB) sidebands caused by the phase difference between AM and FM. The second-order sidebands differ by 1.8–4.2 dB at levels near −40 dB, which is attributable to second-order nonlinear terms that are absent from the linearized power and chirp. With a 1 ns lag window, the numerical PWVD, Fig. 9(b), resolves the lines at half-integer multiples of $f_m$ and coincides with Eq. (24), Fig. 9(c), to within 0.6% of its peak. The odd half-integer lines alternate in sign at $f_m$ and the integer lines beat at $2f_m$, as predicted.

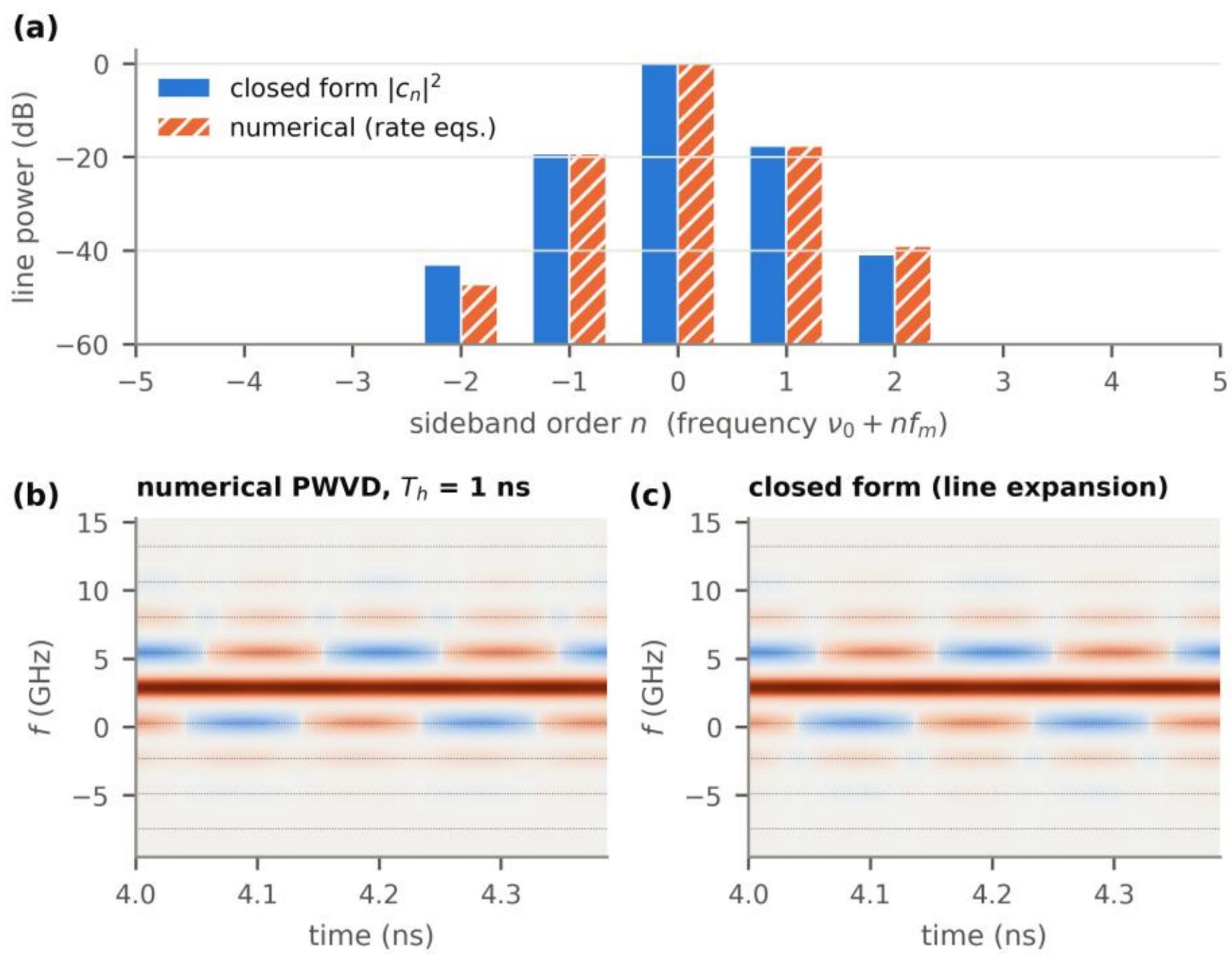

**Fig. 9.** Line structure for $f_m = 5.17$ GHz, $I_m = 0.1I_{th}$, $I_b = 2.5I_{th}$. (a) Powers of the carrier and sidebands: closed form $|c_n|^2$ from Eq. (22) (solid bars) and numerical field (hatched bars). (b) Numerical PWVD with $T_h = 1$ ns and (c) closed-form line expansion, Eq. (24). Dotted horizontal lines mark $\nu_0 + jf_m/2$.

Under large-signal modulation ($I_m = 1.3I_{th}$, Fig. 10), the laser is driven close to threshold on each cycle. At 1 GHz the power varies between 0.79 and 19.7 mW with second and third harmonics at −25.6 and −24.0 dBc; the chirp, between −2.2 and +9.4 GHz, is a distorted replica of the power with a relaxation-oscillation burst at each turn-on, and the ratio of the standard deviations of the adiabatic and transient components is 0.76. At 5.17 GHz the response becomes a gain-switched pulse train with peaks of 51 mW, harmonics at −3.8 and −8.5 dBc, and a sawtooth-like chirp from −15.2 to +33.8 GHz dominated by transient chirp (standard-deviation ratio 0.26). The effective FM indices, 5.8 and 4.7, are 1.6–2 times the small-signal extrapolations (2.9 and 3.0). The time–frequency representation displays harmonic distortion and nonlinear FM together, as a curved and asymmetric ridge whose local slope is the chirp rate.

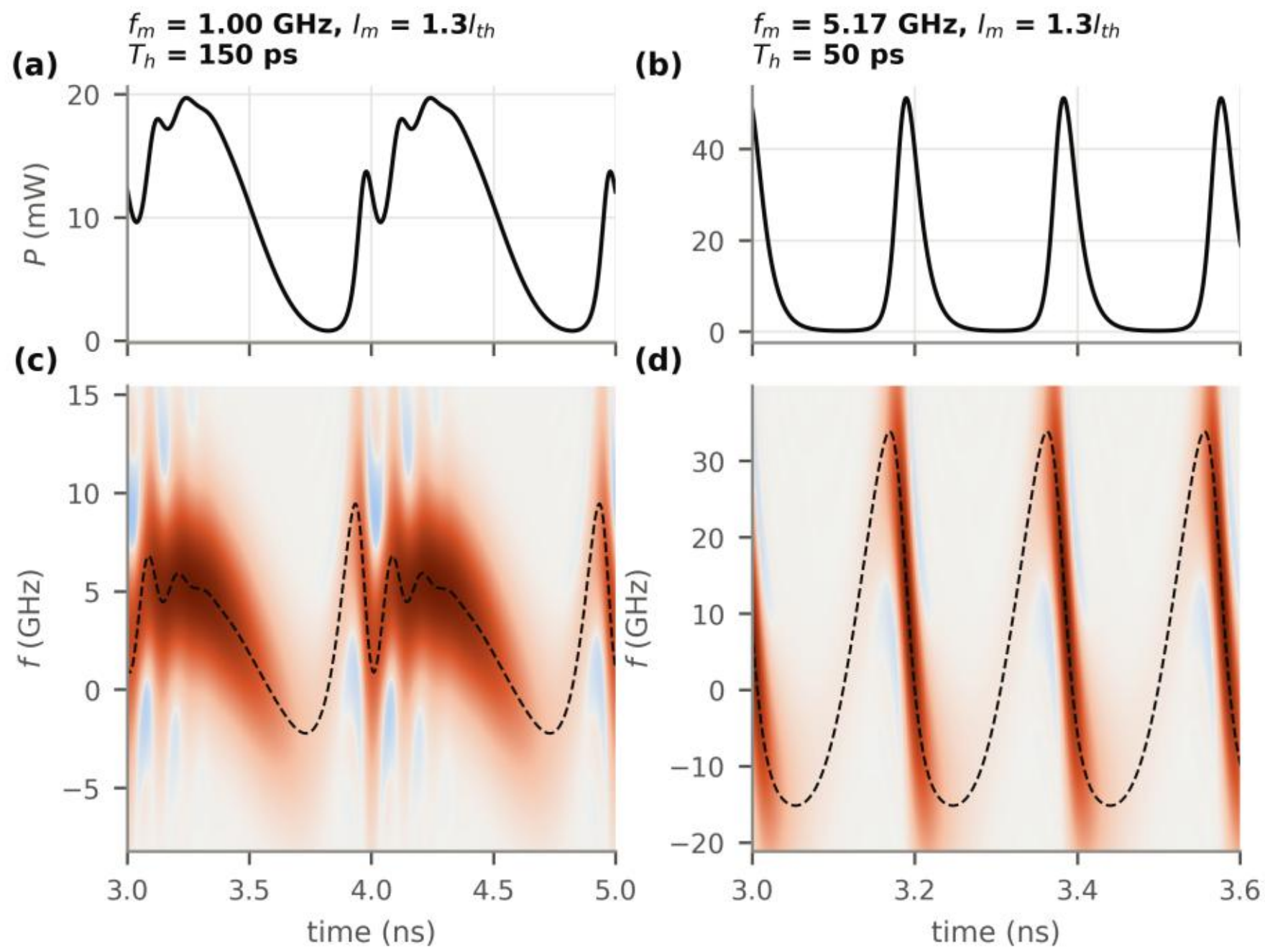

**Fig. 10.** Large-signal sinusoidal modulation ($I_b = 2.5I_{th}$, $I_m = 1.3I_{th}$). (a), (b) Output power and (c), (d) PWVD with the numerical chirp (dashed) for $f_m = 1$ GHz ($T_h = 150$ ps) and $f_m = 5.17$ GHz ($T_h = 50$ ps).

### *9.5 Two-tone modulation, intermodulation and SFDR*

Figure 11 shows the time–frequency representation of a moderate two-tone drive ($f_1 = 2$ GHz, $f_2 = 2.25$ GHz, $I_m = 0.5I_{th}$ per tone on a $2.5I_{th}$ bias). The PWVD displays the envelope beat at $f_2 - f_1$ as a periodic swelling of the frequency excursion, with the chirp ranging from $-1.9$ to +10.1 GHz. Near the envelope maxima, relaxation-oscillation bursts appear at each steep rise of the optical power; these distort the ridge and generate the intermodulation products. Near the envelope minima, the ridge collapses to a nearly constant frequency.

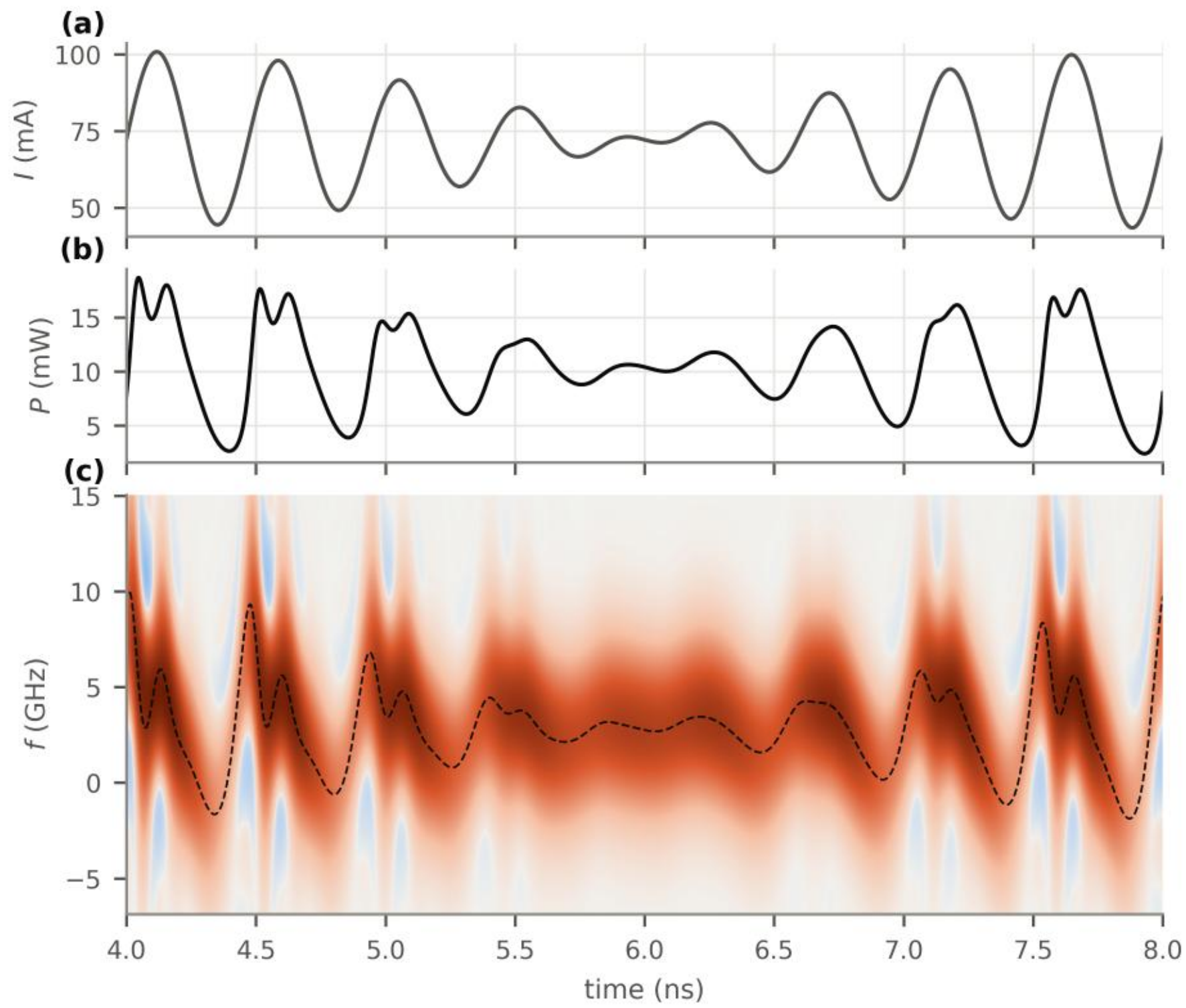

**Fig. 11.** Two-tone modulation ($f_1 = 2$ GHz, $f_2 = 2.25$ GHz, $I_m = 0.5I_{th}$ per tone, $I_b = 2.5I_{th}$) over one beat period. (a) Current; (b) output power; (c) PWVD ($T_h = 150$ ps) with the chirp (dashed).

The Volterra kernels, Eqs. (28)–(33), were compared with lock-in analysis of the numerically integrated two-tone response ($f_1 = 2$ GHz, $f_2 = 2.05$ GHz, $I_b = 2.5I_{th}$). For $I_m = 0.02I_{th}$ all products ($f_1$, $f_1 + f_2$, $f_2 - f_1$, $2f_1$ and $2f_1 - f_2$) agree within 0.11% in amplitude, in both power and chirp. For $I_m = 0.1I_{th}$ the deviations grow to 1.4% ($f_1 + f_2$) and 2.6% ($2f_1 - f_2$), as expected from the neglected fourth- and fifth-order terms. Figure 12 shows the RF output power per tone of the fundamental, IMD2 ($f_1 + f_2$) and IMD3 ($2f_1 - f_2$) products at 2 and 5 GHz. The kernel predictions (lines) follow the slopes 1, 2 and 3, and agree with the full numerical solutions (markers) within 0.11 dB up to −10 dBm input power per tone. At 0 dBm per tone, the numerical IMD3 exceeds the third-order prediction by 1.1 dB (2 GHz) and 0.8 dB (5 GHz), marking the onset of higher-order distortion. At 2 GHz and $2.5I_{th}$, the link gain is −13.6 dB, OIP2 = 12.9 dBm and OIP3 = 17.4 dBm. The noise floor at $2f_1 - f_2$ is −147.1 dBm/Hz, dominated by RIN (−147.4 dBm/Hz, against −158.8 dBm/Hz of detector shot noise and −174.0 dBm/Hz of thermal noise). The resulting dynamic ranges are $SFDR_2$ = 75.7 dB·$Hz^{1/2}$ and $SFDR_3$ = 109.7 dB·$Hz^{2/3}$. At 5 GHz, closer to $f_R$, OIP3 falls to 11.6 dBm and the RIN-dominated floor rises to −134.6 dBm/Hz, so that $SFDR_3$ = 97.4 dB·$Hz^{2/3}$ and $SFDR_2$ = 71.3 dB·$Hz^{1/2}$. In multi-octave systems, in which $f_1 + f_2$ falls in band, the link is therefore limited by second-order distortion; $SFDR_3$ is the relevant figure for sub-octave systems. The difference-frequency product $f_2 - f_1$ is negligible here (OIP2 ≈ 82 dBm).

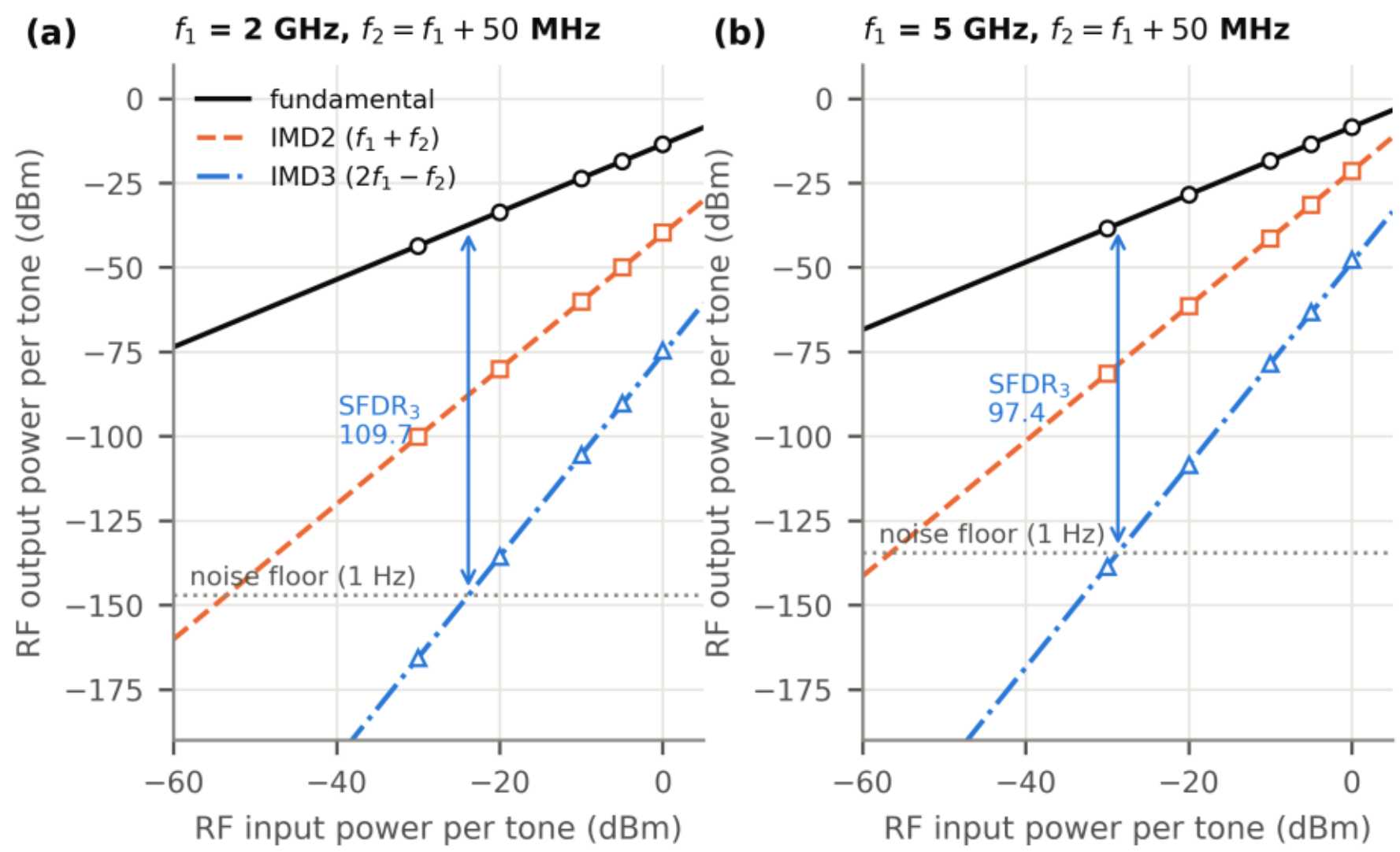

**Fig. 12.** RF output power per tone of the fundamental (solid), IMD2 at $f_1 + f_2$ (dashed) and IMD3 at $2f_1 - f_2$ (dash-dotted) versus RF input power per tone for (a) $f_1 = 2$ GHz and (b) $f_1 = 5$ GHz ($f_2 = f_1 + 50$ MHz, $I_b = 2.5I_{th}$, link of Table 3). Lines: Volterra kernels, Eqs. (28)–(33); markers: full numerical integration. The dotted line is the noise floor in 1 Hz at $2f_1 - f_2$ and the arrow marks SFDR$_3$.

Figure 13 and Table 7 summarize the dependence on frequency, bias and laser. The closed-form RIN, Eq. (37), agrees with a stochastic (Euler–Maruyama) integration of the Langevin rate equations within 0.7 dB between 1 and 11.5 GHz, Fig. 13(a). OIP3 decreases with frequency but is not monotonic: it shows local minima at $f_1 \approx f_R/2$ (1.9, 3.3 and 4.5 GHz at 1.5, 2.5 and $3.5I_{th}$), where the second harmonic generated inside the third-order kernel [through $\mathbf{H}_2(\omega_1, \omega_1)$ in Eq. (31)] is resonant, and at $f_1 \approx f_R$ (3.9, 6.8 and 8.9 GHz) [37]. SFDR$_2$ has a pronounced minimum when the second-order product falls on the resonance, $f_1 + f_2 \approx f_R$ ($f_1 \approx 1.9$, 3.3 and 4.2 GHz), where both the IMD2 product and the noise floor at $f_1 + f_2$ are resonantly enhanced. SFDR$_3$ reaches its minimum near $f_1 \approx f_R$ (82.7, 92.2 and 96.8 dB·Hz$^{2/3}$ at 3.8, 6.6 and 8.5 GHz). Below about 3 GHz both SFDRs improve with bias, because $f_R$ and the damping increase and the RIN falls; at higher frequencies the curves cross, because the resonant features move to higher frequency with bias. At 2 GHz, raising the bias from $1.5I_{th}$ to $3.5I_{th}$ improves SFDR$_3$ from 94.8 to 115.6 dB·Hz$^{2/3}$ and SFDR$_2$ from 57.7 to 82.8 dB·Hz$^{1/2}$.

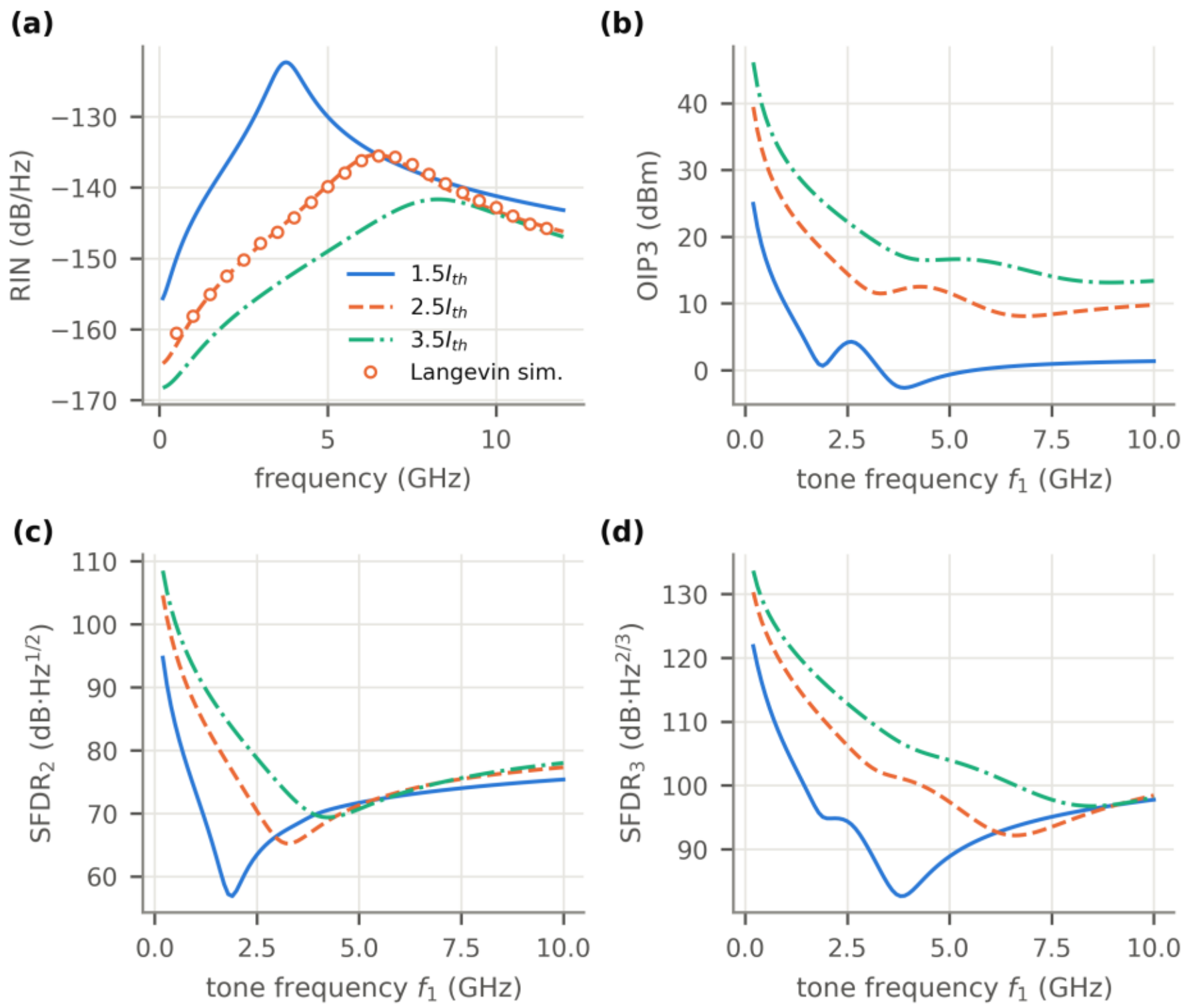

**Fig. 13.** Noise and dynamic range of the DFB-1550 link at $1.5I_{th}$ (solid), $2.5I_{th}$ (dashed) and $3.5I_{th}$ (dash-dotted). (a) RIN from Eq. (37) and from stochastic integration at $2.5I_{th}$ (circles); (b) OIP3; (c) SFDR$_2$ and (d) SFDR$_3$ in a 1 Hz bandwidth versus the frequency of the first tone ($f_2 = f_1 + 50$ MHz, $f_1$ up to 10 GHz).

**Table 7. Two-tone link metrics ($f_2 = f_1 + 50$ MHz, link of Table 3). $\mathcal{N}_3$ is the noise density at $2f_1 - f_2$; SFDR$_2$ uses the noise density at $f_1 + f_2$.**

| Laser, bias | $f_1$ [GHz] | Gain [dB] | OIP2 [dBm] | OIP3 [dBm] | RIN [dB/Hz] | $\mathcal{N}_3$ [dBm/Hz] | SFDR$_2$ [dB·Hz$^{1/2}$] | SFDR$_3$ [dB·Hz$^{2/3}$] |
|---|---|---|---|---|---|---|---|---|
| DFB-1550, $1.5I_{th}$ | 1 | −13.8 | 6.4 | 9.7 | −145.0 | −149.0 | 73.4 | 105.8 |
| DFB-1550, $1.5I_{th}$ | 2 | −11.7 | −12.4 | 1.1 | −137.0 | −141.2 | 57.7 | 94.8 |
| DFB-1550, $2.5I_{th}$ | 1 | −14.2 | 27.6 | 24.7 | −158.7 | −152.2 | 87.1 | 118.0 |
| DFB-1550, $2.5I_{th}$ | 2 | −13.6 | 12.9 | 17.4 | −152.8 | −147.1 | 75.7 | 109.7 |
| DFB-1550, $2.5I_{th}$ | 5 | −8.5 | 4.5 | 11.6 | −139.9 | −134.6 | 71.3 | 97.4 |
| DFB-1550, $3.5I_{th}$ | 1 | −14.3 | 36.8 | 31.5 | −164.3 | −152.4 | 92.6 | 122.6 |
| DFB-1550, $3.5I_{th}$ | 2 | −13.9 | 23.5 | 24.7 | −159.3 | −148.7 | 82.8 | 115.6 |
| DFB-1550, $3.5I_{th}$ | 5 | −11.3 | 7.4 | 16.6 | −149.2 | −139.3 | 70.7 | 104.0 |
| DFB-1310, $2.5I_{th}$ | 1 | −12.8 | 32.1 | 26.6 | −159.7 | −153.5 | 90.6 | 120.1 |

| Laser, bias | $f_1$ [GHz] | Gain [dB] | OIP2 [dBm] | OIP3 [dBm] | RIN [dB/Hz] | $\mathcal{N}_3$ [dBm/Hz] | $SFDR_2$ [dB·Hz$^{1/2}$] | $SFDR_3$ [dB·Hz$^{2/3}$] |
|---|---|---|---|---|---|---|---|---|
| DFB-1310, $2.5I_{th}$ | 2 | −12.4 | 18.7 | 19.9 | −154.7 | −149.4 | 80.5 | 112.9 |
| VCSEL-850, $2.5I_{th}$ | 1 | −7.0 | −6.5 | −8.6 | −139.4 | −165.5 | 76.8 | 104.6 |
| VCSEL-850, $2.5I_{th}$ | 2 | −6.2 | −22.2 | −16.3 | −133.5 | −160.5 | 64.3 | 96.1 |

RIN is evaluated at $2f_1 - f_2$. VCSEL link with $\mathcal{R} = 0.6$ A/W.

Among the three lasers, the DFB-1310 laser gives the largest dynamic range at 1–2 GHz ($SFDR_3$ = 120.1 and 112.9 dB·Hz$^{2/3}$), a consequence of its higher relaxation frequency. The VCSEL has the highest link gain (−7.0 dB, with $\mathcal{R} = 0.6$ A/W) owing to its high slope efficiency, but low intercept points owing to its sub-milliwatt output. Its small photocurrent (0.19 mA) keeps the absolute noise floor low (−165.5 dBm/Hz) despite a RIN of −139 dB/Hz, and its $SFDR_3$ is 104.6 dB·Hz$^{2/3}$ at 1 GHz. The kernels also quantify the intermodulation of the frequency modulation. At 2 GHz and −10 dBm per tone ($I_m = 2$ mA), the IMD3 of the power is −82.0 dB relative to the fundamental, whereas the IMD3 of the chirp is −64.3 dB; the corresponding IMD2 ratios are −36.5 and −25.1 dB. The frequency modulation is thus markedly more nonlinear than the intensity modulation. In dispersive links, dispersion converts frequency modulation into intensity modulation [48], and this FM distortion can then dominate the detected intermodulation. This effect is not included in the back-to-back SFDR values above. Because the model excludes static nonlinearities of the light–current characteristic (leakage and thermal effects), spatial hole burning, clipping and dispersion, the SFDR values should be regarded as intrinsic upper bounds set by the carrier–photon dynamics.

### *9.6 Moment accuracy, bandwidth budget and parameter extraction*

Figure 14(a) quantifies Eq. (44) (CWD without lag window) on the 16-bit NRZ record of Fig. 7(d) through the median and 99th percentile of $|\langle f \rangle_t - \phi'/2\pi|$. The PWVD with $T_h = 20$ ps reproduces the chirp with a median error of 1.6 Hz and a 99th-percentile error of 0.30 MHz, the residual reflecting the lag-domain discretization of the derivative. The CWD with $\sigma = 3$ reaches 25 kHz and 5.1 MHz. The discrete WVD, although exact in continuous time, has a 99th-percentile error of 49 GHz, located in the low-power intervals after "1"→"0" transitions, which is the non-locality discussed in Section 7. Kernels with time smoothing return a smoothed chirp with 99th-percentile errors of 6.2 GHz (SPWVD) and 11.2 GHz (spectrogram), which are errors of principle and not of discretization.

The rms-bandwidth budget, Eq. (47), is tested in Fig. 14(d) on a finite-energy pulse (RC-filtered, $0.9 \rightarrow 3.5I_{th}$) for three values of $\varepsilon$. Cohen's identity, Eq. (45), agrees with the rms width of the FFT spectrum to better than 0.05%. The budget of Eqs. (47) and (48) underestimates the spectral width by 0.5%, 0.9% and 1.9% for $\varepsilon = 10^{-24}$, $1.5 \times 10^{-23}$ and $5 \times 10^{-23}$ m³; with $\beta = 10^{-6}$ the deviation falls below 0.08%, which identifies the spontaneous-emission term, large in the dark tails of the pulse, as the origin of the residual. The weak-compression form $(1 + \alpha^2)B_{TL}^2 + \sigma_{ad}^2$ underestimates it by 0.8%, 3.9% and 6.3%. The transient term dominates the budget: $B_{tr}^2$ accounts for about 86% of $B^2$ at the nominal compression, and the adiabatic variance becomes significant only at strong compression.

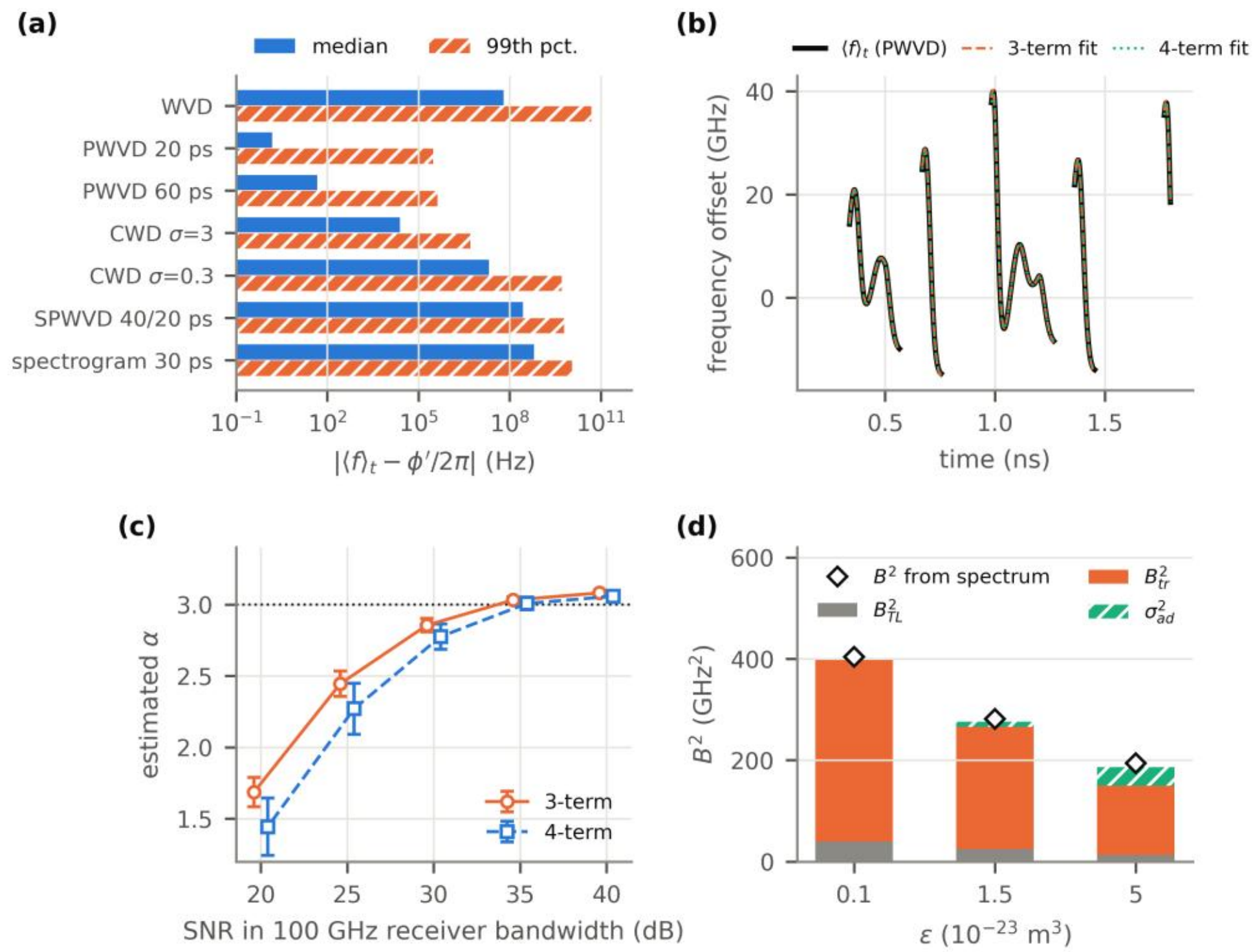

**Fig. 14.** Accuracy and parameter extraction. (a) Median (solid) and 99th-percentile (hatched) error of the first conditional moment with respect to the rate-equation chirp for the 10 Gb/s NRZ record; (b) first conditional moment of the PWVD ($T_h = 20$ ps) and the three-term and four-term fits of Eq. (49); (c) $\alpha$ estimated with the three-term (circles) and four-term (squares) regressions versus SNR in a 100 GHz receiver bandwidth (mean ± standard deviation over 20 realizations; dotted line: true value); (d) rms-bandwidth budget of Eqs. (47) and (48) for an RC-filtered pulse from $0.9I_{th}$, compared with the rms width of the FFT spectrum (diamonds).

The results of the parameter extraction of Eq. (49) are collected in Table 8. With the PWVD or the CWD and noise-free fields, the four-term estimator recovers $\alpha = 3.000$ within 0.1% and $\tau_p$ within 0.8% for all four drive conditions and both compression values. It recovers the adiabatic coefficient within 1.0%, the residual being attributable to the neglected spontaneous-emission term. The three-term Koch–Linke estimator overestimates $\alpha$ by 1.2–6.6%, the error growing with the peak power and with $\varepsilon$, because it attributes the $(1 + \varepsilon S)$ factor of the transient term to $\alpha$. The spectrogram, whose centroid is not the chirp, overestimates $\alpha$ by up to 17% and yields photon lifetimes between 1.7 and 3.3 ps. Sinusoidal large-signal modulation and RC-filtered NRZ patterns are equally suitable excitations; the fit residuals of the four-term estimator are 6–14 MHz, compared with chirp excursions of tens of GHz. These noise-free results establish the consistency of the estimators with the model; their practical accuracy is limited by measurement noise and bandwidth, examined next.

**Table 8. Parameter extraction from the first conditional moment for the DFB-1550 laser ($\varepsilon = 1.5 \times 10^{-23}$ m³; true values $\alpha = 3$, $\alpha\kappa/4\pi = 0.2794$ GHz/mW, $\tau_p = 2$ ps). "3-term": Koch–Linke regression; "4-term": Eq. (49). Values for $\varepsilon = 5 \times 10^{-23}$ m³ are given in the text.**

| Drive | Distribution | $\alpha$ (3-term) | $\alpha$ (4-term) | $\alpha\kappa/4\pi$ (4-term) [GHz/mW] | $\tau_p$ (4-term) [ps] | Residual (4-term) [MHz] |
|---|---|---|---|---|---|---|
| RC-NRZ, 10 Gb/s | PWVD, $T_h = 20$ ps | 3.101 | 3.001 | 0.2795 | 2.01 | 6 |

| Drive | Distribution | $\alpha$ (3-term) | $\alpha$ (4-term) | $\alpha\kappa/4\pi$ (4-term) [GHz/mW] | $\tau_p$ (4-term) [ps] | Residual (4-term) [MHz] |
|---|---|---|---|---|---|---|
| RC-NRZ, 10 Gb/s | spectrogram, $T_w =$ 30 ps | 3.319 | 3.243 | 0.2864 | 2.09 | 1080 |
| single RC pulse | PWVD, $T_h = 20$ ps | 3.065 | 3.000 | 0.2804 | 2.01 | 9 |
| single RC pulse | spectrogram, $T_w =$ 30 ps | 3.132 | 3.037 | 0.3074 | 3.28 | 395 |
| sine, 1 GHz, $1.3I_{th}$ | PWVD, $T_h = 20$ ps | 3.037 | 2.999 | 0.2818 | 2.01 | 12 |
| sine, 1 GHz, $1.3I_{th}$ | spectrogram, $T_w =$ 30 ps | 3.081 | 3.028 | 0.2796 | 3.00 | 34 |
| sine, 5.17 GHz, $1.3I_{th}$ | PWVD, $T_h = 20$ ps | 3.097 | 3.001 | 0.2794 | 2.01 | 9 |
| sine, 5.17 GHz, $1.3I_{th}$ | spectrogram, $T_w =$ 30 ps | 3.298 | 3.244 | 0.2588 | 1.73 | 655 |

CWD ($\sigma = 3$, 200 ps lag window) results coincide with the PWVD results to the digits shown.

For $\varepsilon = 5 \times 10^{-23}$ m³ (true $\alpha\kappa/4\pi = 0.9314$ GHz/mW), the four-term PWVD estimator gives $\alpha = 3.000$–$3.001$, $\alpha\kappa/4\pi = 0.9326$–$0.9345$ GHz/mW and $\tau_p = 1.99$–$2.00$ ps over the four drives, whereas the three-term estimator gives $\alpha = 3.10$–$3.20$.

The robustness of the estimators is summarized in Table 9 and Fig. 14(c). A finite receiver bandwidth smooths both the power and the chirp: without noise, the four-term estimate of $\alpha$ is within 0.9% of the true value at $B_{rx} = 200$ GHz and within 3.4% at 100 GHz, but the adiabatic coefficient is underestimated by 2.4% and 9.7%, respectively, and the photon lifetime, which rests on the small correction term $a_3$, is unusable below $B_{rx} \approx 200$ GHz. Additive noise biases $\alpha$ downward, because the noisy regressor $d\ln P/dt$ attenuates the fitted slope (errors-in-variables bias): with $B_{rx} = 100$ GHz, the three-term estimate is $3.03 \pm 0.02$ at 35 dB SNR, $2.85 \pm 0.05$ at 30 dB and $1.69 \pm 0.10$ at 20 dB, and the four-term estimate is less robust, with about twice the spread. The adiabatic coefficient is less sensitive to noise (0.251–0.256 GHz/mW for SNR between 30 and 40 dB, i.e., dominated by the 9.7% bandwidth bias). We also verified that, in the presence of noise, the lag-windowed centroid coincides with the numerical derivative of the unwrapped phase (0.8% relative rms difference at 30 dB SNR), as implied by Eq. (44): a lag window does not smooth the first moment, and noise reduction requires time smoothing, i.e., kernels with $\Phi(\nu, 0) \neq 1$, at the price of the bias quantified in Fig. 14(a). In practice, reliable extraction of $\alpha$ requires an SNR of about 35 dB in a receiver bandwidth several times larger than the chirp excursion; under these conditions the four-term form is preferable because it removes the compression bias of the three-term form.

**Table 9. Sensitivity of the moment-based estimators to the receiver bandwidth (noise-free) and to additive noise ($B_{rx} = 100$ GHz), 10 Gb/s RC-NRZ record, PWVD with $T_h = 20$ ps. True values: $\alpha = 3$, $\alpha\kappa/4\pi =$ 0.2794 GHz/mW, $\tau_p = 2$ ps. Noise results: mean ± standard deviation over 20 realizations.**

| Condition | $\alpha$ (3-term) | $\alpha$ (4-term) | $\alpha\kappa/4\pi$ [GHz/mW] | $\tau_p$ (4-term) [ps] |
|---|---|---|---|---|
| no filter, no noise | 3.101 | 3.001 | 0.2795 | 2.01 |
| $B_{rx} = 200$ GHz | 3.105 | 3.027 | 0.2726 | 1.62 |
| $B_{rx} = 100$ GHz | 3.109 | 3.103 | 0.2523 | 0.15 |
| $B_{rx} = 50$ GHz | 3.015 | 3.282 | 0.1825 | −9.2 |
| SNR 40 dB | 3.081 ± 0.009 | 3.058 ± 0.022 | 0.2519 ± 0.0032 | 0.5 ± 0.3 |
| SNR 35 dB | 3.031 ± 0.022 | 3.007 ± 0.043 | 0.2515 ± 0.0043 | 0.6 ± 0.6 |
| SNR 30 dB | 2.854 ± 0.047 | 2.775 ± 0.087 | 0.2560 ± 0.0091 | 1.8 ± 1.0 |

| Condition | $\alpha$ (3-term) | $\alpha$ (4-term) | $\alpha\kappa/4\pi$ [GHz/mW] | $\tau_p$ (4-term) [ps] |
|---|---|---|---|---|
| SNR 25 dB | 2.445 ± 0.089 | 2.271 ± 0.178 | 0.2525 ± 0.0092 | 4.2 ± 2.4 |
| SNR 20 dB | 1.688 ± 0.102 | 1.447 ± 0.199 | 0.2443 ± 0.0245 | 6.3 ± 3.2 |

### *9.7 Kernel selection and cross-terms*

The choice of kernel is dictated by the geometry of the interference terms in the ambiguity plane [19,47], and this geometry depends on the modulation format. Figure 15 considers two RC-filtered pulses separated by 600 ps. The WVD, Fig. 15(a), contains a strong interference pattern midway between the pulses, oscillating along frequency with period $1/(600\text{ ps}) \approx 1.7$ GHz. In the ambiguity plane, Fig. 15(c), the corresponding cross-terms are located at $\tau = \pm 600$ ps and close to $\nu = 0$, because the two pulses have similar spectra. The CWD kernel is unity along the whole $\nu = 0$ axis and therefore cannot suppress them, which also explains the horizontal striping of the CWD in long records containing steady segments. A lag window of a few tens of picoseconds removes them, Fig. 15(b), without affecting the auto-terms, whose extent in $\tau$ is set by the pulse duration. For pulse trains and NRZ data, the PWVD with $T_h$ between the transition time and the bit period is therefore the natural choice. For analog modulation, $T_h$ should be chosen according to Section 5.2. The spectrogram is useful when a non-negative display is required, but not for quantitative chirp analysis.

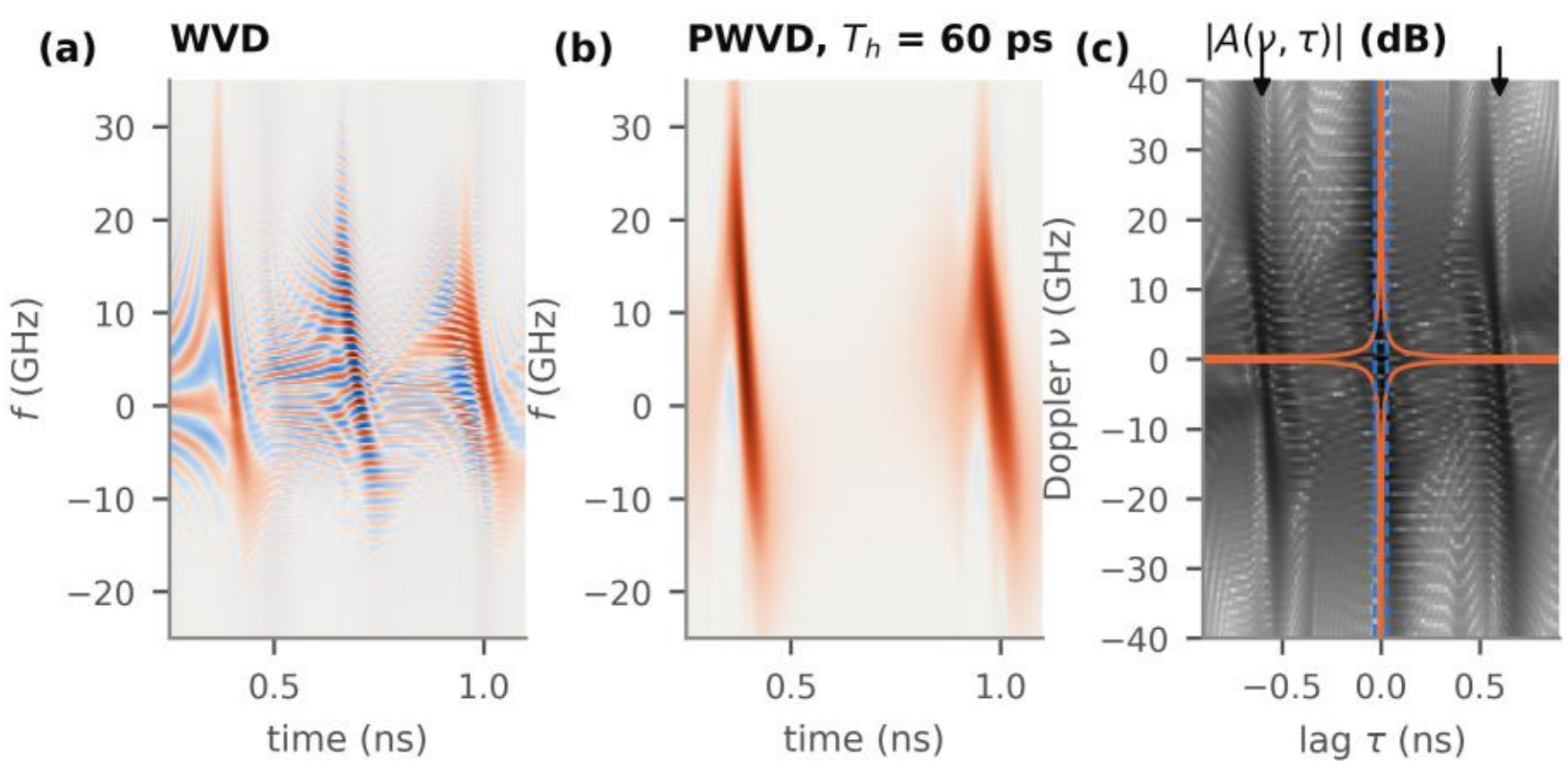

**Fig. 15.** Cross-terms and kernel design for two RC-filtered pulses 600 ps apart. (a) WVD; (b) PWVD with $T_h = 60$ ps; (c) magnitude of the ambiguity function (gray scale, dB), with the 50% contour of the Choi–Williams kernel ($\sigma = 1$, orange) and the lag window $|\tau| \leq 30$ ps (blue dashed). Arrows mark the pulse–pulse cross-terms at $\tau = \pm 600$ ps.

### *9.8 Comparison of laser types*

Finally, the three lasers of Table 1 are driven by the same normalized 10 Gb/s RC-filtered NRZ pattern ($1.5 \rightarrow 5I_{th}$), Fig. 16, and their small-signal and extracted parameters are compared in Table 10. The 1310 nm DFB laser has a higher relaxation frequency (8.6 GHz at $2.5I_{th}$ versus 6.5 GHz) and a smaller $\alpha$. Its time–frequency signature shows faster and more strongly underdamped ringing, but transient strokes of comparable span (42 GHz peak-to-peak versus 41 GHz), because its sharper spikes compensate for the smaller $\alpha$. The VCSEL, with sub-milliwatt power and the largest $\alpha$, produces the widest transient excursions (54 GHz peak-to-peak) and the largest adiabatic coefficient per unit power (1.59 GHz/mW), a consequence of its small $P_S$ [Eq. (7)]. The four-term estimator recovers $\alpha$ within 0.1% for the three lasers. The photon lifetime is recovered within 0.5% for the DFB-1550 laser and within 2.8% and 4.4% for the DFB-1310 laser and the VCSEL. These errors are essentially unchanged when spontaneous emission is removed ($\beta = 10^{-7}$: 1.645 and 2.627 ps), so they are not due to the $\beta$ term. They reflect the conditioning of the small correction $a_3$, whose regressor is weighted by $\varepsilon S$ (smaller in these two lasers than in the DFB-1550 laser at the same normalized drive). They also reflect the finite-difference evaluation of $d\ln P/dt$ on the very sharp spikes of the VCSEL. Removing spontaneous emission does reduce the VCSEL adiabatic-coefficient error from 2.3% to 0.4%.

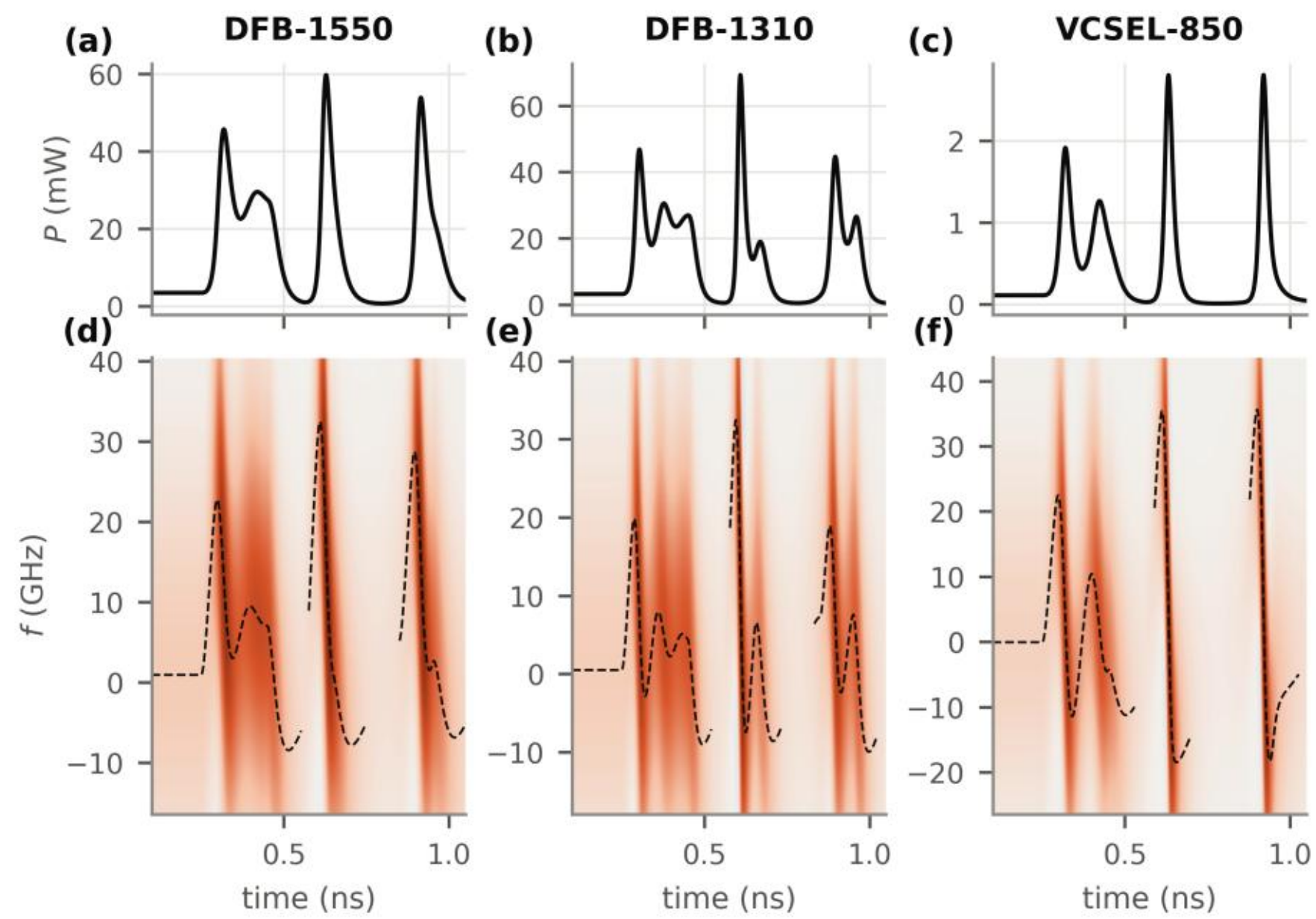

**Fig. 16.** Three lasers under the same normalized RC-filtered NRZ drive (10 Gb/s, $\tau_{RC} = 15$ ps, $1.5 \to 5I_{th}$). (a)–(c) Output power and (d)–(f) PWVD ($T_h = 30$ ps) with the chirp (dashed) for DFB-1550, DFB-1310 and VCSEL-850.

**Table 10. Small-signal quantities at $2.5I_{th}$ and parameters extracted with the three- and four-term estimators (PWVD, $T_h = 20$ ps) from the noise-free records of Fig. 16.**

| Quantity | DFB-1550 | DFB-1310 | VCSEL-850 |
|---|---|---|---|
| $I_{th}$ [mA] | 28.9 | 22.8 | 0.293 |
| $\overline{P}$ at $2.5I_{th}$ [mW] | 10.31 | 9.65 | 0.321 |
| $f_R$ at $2.5I_{th}$ / $5I_{th}$ [GHz] | 6.46 / 10.34 | 8.56 / 13.81 | 6.08 / 9.89 |
| $\gamma/2\pi$ at $2.5I_{th}$ [GHz] | 2.56 | 2.35 | 1.07 |
| $f_c$ at $2.5I_{th}$ [GHz] | 1.88 | 1.39 | 0.33 |
| $\alpha$: true / 3-term / 4-term | 3.00 / 3.115 / 3.001 | 2.20 / 2.260 / 2.200 | 3.50 / 3.540 / 3.498 |
| $\alpha\kappa/4\pi$: true / 4-term [GHz/mW] | 0.279 / 0.280 | 0.160 / 0.160 | 1.589 / 1.625 |
| $\tau_p$: true / 4-term [ps] | 2.00 / 2.009 | 1.60 / 1.645 | 2.50 / 2.609 |
| fit residual (4-term) [MHz] | 8.5 | 7.5 | 29 |

## 10. Discussion

The results establish three properties that make time–frequency distributions a quantitative tool for laser dynamics rather than a visualization aid. First, for a suitable kernel the distribution contains the laser chirp exactly, as its first conditional moment, together with the optical power (time marginal) and, for the WVD and CWD, the optical spectrum (frequency marginal). A single object therefore unifies the time-resolved chirp measurements used to characterize DMLs [43] and the spectral measurements used to characterize their sidebands [12,13]. Second, the $\alpha$-factor acquires a geometric meaning, since it relates the displacement of the centroid to the instantaneous bandwidth [Eq. (46)], and it controls the global rms bandwidth through the simple budget of Eq. (47). Third, the Koch–Linke analysis can be carried out on the moments of the distribution. This is equivalent to an analysis of the time-resolved chirp, since the centroid equals $\phi'/2\pi$. Two points matter in practice. Retaining the compression factor in the transient term removes a bias of up to 7% in $\alpha$ under large-signal conditions. Moments of time-smoothing distributions, which include the spectrograms underlying FROG-type measurements, return a smoothed chirp and bias $\alpha$ by up to 17% unless the smoothing is accounted for. The robustness study shows that an SNR of about 35 dB and a receiver bandwidth well above the chirp excursion are needed for accurate $\alpha$, and that the photon-lifetime estimate is of theoretical rather than practical interest.

The closed-form solutions are useful beyond the time–frequency context. Equations (9)–(13) give the IM response and the CPR with compression and spontaneous emission included exactly. Equations (15)–(17) give the complete small-signal pulse response for a realistic, RC-limited driver, and Eq. (18) the turn-on delay for such a driver. The line expansion of Eq. (24) explains why the chirp of analog links with $\beta_{FM} \ll 1$ does not appear as a conspicuous trajectory, although it is carried exactly by the centroid, and shows how the lag window should be chosen.

The two-tone analysis extends the closed-form treatment to the figures of merit of analog photonic links. The Volterra kernels give the intercept points exactly in the small-signal limit, including gain compression and spontaneous emission. Combined with the Langevin RIN, they give SFDR maps as a function of bias and frequency without large-signal simulations; the resonant minima of $\text{SFDR}_2$ at $f_1 + f_2 \approx f_R$ and of $\text{SFDR}_3$ near $f_R$ (Fig. 13) follow directly from the poles of the kernels. Several limitations must be acknowledged. The model is single-mode; the time-domain simulations are deterministic, and noise enters only through the small-signal Langevin RIN of Eq. (37). Thermal chirp, carrier transport, spatial hole burning and, for VCSELs, transverse-mode competition are not included, and the laser parameters are representative values rather than measurements of specific devices. The SFDR values refer to a back-to-back link with a matched laser input and a photodiode delivering its full current into the load; they depend on the assumed $n_{sp}$ and pump-noise model and exclude fiber dispersion, which converts the stronger FM intermodulation into intensity distortion. The chirp is undefined when the laser is below threshold and the field is dominated by spontaneous emission; the corresponding intervals were excluded from all quantitative analyses. On the experimental side, the framework requires the complex field $E(t)$. This can be obtained with a coherent receiver or linear optical sampling [49], or reconstructed from spectrographic measurements [24–27]. The alias-free condition of Eq. (50) is a computational constraint: a Nyquist-sampled complex field can be interpolated by a factor of two before the distribution is computed. Extending the time–frequency analysis to noisy fields (where the ensemble-averaged Wigner distribution is related to the field coherence), fiber dispersion (which acts on the Wigner distribution as a shear along time), and measured fields is the natural continuation of this work.

## 11. Conclusion

We have presented a time–frequency framework for the dynamics of directly modulated semiconductor lasers based on Cohen-class representations of the field predicted by the rate equations. We derived an exact decomposition of the chirp, closed-form small-signal solutions for RC-filtered square-pulse and sinusoidal modulation, and a line expansion of the Wigner distribution of the sinusoidally modulated field, and validated them against numerical integration to within the numerical tolerance or to the leading nonlinear correction. Kernels with $\Phi(\nu, 0) = 1$ and $\partial_\tau \Phi(\nu, 0) = 0$ return the chirp exactly; the transient chirp is $\alpha(1 + \varepsilon S)$ times the instantaneous bandwidth; and the rms bandwidth of a chirped pulse obeys an exact budget that reduces to $B^2 = (1 + \alpha^2) B_{TL}^2 + \sigma_{ad}^2$ for weak compression. A four-term Koch–Linke regression on the moments recovers the linewidth-enhancement factor within 0.1% from noise-free fields of three different lasers under digital and analog excitation, whereas spectrogram moments are biased. Accurate extraction from measured fields requires an SNR of about 35 dB and a receiver bandwidth well above the chirp excursion. For analog links, third-order Volterra kernels of the rate equations and a Langevin RIN model give the intercept points and the SFDRs in 1 Hz. The kernels agree with full simulations within 0.11 dB. For the reference laser at $2.5I_{th}$ they give $SFDR_2$ = 75.7 dB·$Hz^{1/2}$ and $SFDR_3$ = 109.7 dB·$Hz^{2/3}$ at 2 GHz, with resonant degradation near $f_R$.

**Funding.** [To be completed by the author.]

**Acknowledgment.** [Optional; to be completed by the author.]

**Disclosures.** The author declares no conflicts of interest.

**Data availability.** Data and simulation code underlying the results presented in this paper are not publicly available at this time but may be obtained from the author upon reasonable request.